\documentclass[aps,pra,superscriptaddress,singlecolumn,showpacs,preprintnumbers,amsmath,amssymb,natbib,balancelastpage,longbibliography]{revtex4-2}
\usepackage[dvips]{graphicx}
\usepackage{srcltx}
\usepackage{color}
\usepackage{bm,upgreek}
\usepackage{amsmath,mathtools}
\usepackage{amsfonts}
\usepackage{amssymb}
\usepackage{mathrsfs}
\usepackage{boxedminipage}
\usepackage{framed}
\usepackage{bbm}
\usepackage{natbib}\usepackage[ugly]{nicefrac}
\usepackage{mathtools}
\usepackage{enumitem}
\usepackage{multirow}
\usepackage{diagbox}

\def\del#1{\color{grey}\sout{#1}~\color{black}}

\def\bra#1{{\langle #1|}}                     
\def\ket#1{{|#1\rangle}}                      
\def\braket#1#2{{\langle #1|#2\rangle}}       
\def\ketbra#1#2{{|#1\rangle\langle#2|}}       
\def\matel#1#2#3{{\bra{#1}#2\ket{#3}}}        
\def\mod2#1{{\big|#1\big|^2}}                 
\def\av#1{{\langle#1\rangle}}                 
\def\set#1{{\left\{#1\right\}}}               
\def\-{\!-\!}                               
\def\+{\!+\!}                               
\def\={\;=\;}                               

\def\rr{\vec{r}}
\def\g{\vec{g}}

\def\k{\vec{k}}

\def\V{\op{V}}

\def\Z{\op{Z}}
\def\1{\op{1}}

\def\0{\vec{0}}

\def\d{\mathrm{d}}                          
\def\intd#1{\int\d#1\;}                     
\def\intd2#1{\int\d^2#1\;}                  
\def\intd3#1{\int\d^3#1\;}                  
\def\del{\partial}                          
\def\vec#1{\mathbf{#1}}                     
\def\op#1{\hat{#1}}                         
\def\set#1{\{#1\}}                          
\def\-{\!-\!}                               
\def\+{\!+\!}                               

\def\beq{\begin{equation}}                  
\def\eeq{\end{equation}}                    

\def\F{\mathscr{F}}                         

\makeatletter
\newcommand*{\lineovertext}[1]{$\overline{\hbox{#1}}\m@th$}
\makeatother

\begin{document}


\title{Generalized Zernike Phase-Contrast Imaging}

\author{Christian Dwyer}
\email{christian.dwyer@dectris.com, christian.dwyer@rmit.edu.au}
\affiliation{DECTRIS Ltd., Baden-Daettwil, Switzerland}
\affiliation{Physics, School of Science, RMIT University, Melbourne, Victoria 3001, Australia}

\author{David M. Paganin}
\email{david.paganin@monash.edu}
\affiliation{School of Physics and Astronomy, Monash University, Clayton, Victoria 3800, Australia}


\begin{abstract}
Zernike phase-contrast imaging is unique among imaging techniques in that it enables the upper limit of Fisher information allowed by quantum mechanics. Here we show that, in a departure from an ideal setting, using an incident beam of finite width, and a $\pi/2$ phase plate having a finite cutoff, the technique can still deliver $>95$\% of the quantum limit. We point out that the Zernike method is, in principle, applicable to any incident beam. As an example, we sketch an approximate implementation of the method for an incident speckle beam, and show that it too can deliver $>95$\% of the quantum limit.
\end{abstract}

\maketitle

\section{Introduction}

In previous works \cite{Dwyer2023, DwyerPaganin2024}, we used quantum estimation theory to show that maximal Fisher information regarding the parameters of a weakly-scattering sample demands Zernike phase-contrast imaging, where (1) post-sample optics imposes a phase shift of $\pm\pi/2$ on the unscattered wave (the Zernike phase condition), and (2) the detection process corresponds to recording the intensity distribution in coordinate space. The Zernike phase condition makes the resulting coordinate-space wavefunction real, thereby achieving maximal interference between the scattered and unscattered waves. The claim of optimality refers only to the spatial frequencies which are not blocked or significantly aberrated by a given imaging system. Conventionally, the Zernike phase condition is achieved using plane-wave illumination and a small $\pi/2$ phase plate in Fourier space, ideally encompassing only the unscattered wave. Our proof \cite{Dwyer2023, DwyerPaganin2024} implies that any imaging technique which does not meet the Zernike phase condition is suboptimally sensitive from the perspective of Fisher information, when the sample being imaged is weakly scattering.

In the present work, we generalize our previous work on an ideal Zernike imaging system by considering illumination having a finite width and a phase plate having finite radius (Section~\ref{sec:finite beam}). We find that these departures from ideality can still allow greater than 95\% of the Fisher information to be obtained. We then construct a unitary operator which describes the action of the post-sample optics for the case of arbitrary coherent illumination (Section~\ref{sec:Zernike operator}). The existence a such an operator implies that, in principle, the Zernike phase condition can be obtained for arbitrary illumination. As a special case, we consider Zernike speckle imaging, whereby the incident wave is spatially random (Section~\ref{sec:Zernike speckle imaging}). We conclude with a brief discussion (Section~\ref{sec:discussion}).

\section{Zernike phase-contrast imaging using finite beam and phase plate}
\label{sec:finite beam}

The results presented below are applicable to several types of coherent, monochromatic radiation, such as visible light, x-rays and electrons. However, for concreteness,  from this point onwards we will adopt notation and nomenclature that is specific to transmission electron microscopy (TEM).

We adopt a disc-shaped, unit-amplitude incident wavefunction (which is an image of the condenser aperture)
\beq \psi_0(\rr) = \theta(R-r), \eeq
where $R$ is the radius and $\theta$ is the Heaviside step function. We use the Fourier transform convention 
\beq \tilde f(\k) = \int\d^2\rr\, f(\rr) e^{-2\pi i\k\cdot\rr}, \qquad f(\rr) = \int\d^2\k\, \tilde f(\k) e^{+2\pi i\k\cdot\rr}, \eeq
where $\rr$ labels the two dimensional coordinate space transverse to the optic axis, and $\k$ labels the corresponding Fourier space. The Fourier transform of $\psi_0$ is an Airy disc in Fourier space \cite{BornWolf1999}
\beq\label{eq:psi0k} \tilde\psi_0(\k) = \frac{R J_1(2\pi Rk)}{k} ,\eeq
where $J_1$ is the first-order Bessel function of the first kind. Adopting the weak phase-object approximation, the exit wavefunction is
\beq \psi_\mathrm{exit}(\rr) = \psi_0(\rr) + iV(\rr)\psi_0(\rr),\eeq
where $V(\rr)$ is the sample potential projected along the direction of the optic axis, and the important factor $i$ represents the $\pi/2$ phase shift on scattering. A Zernike phase plate at the origin of Fourier space is described by the function
\beq \tilde Z(\k) = 1 + (i-1)\theta(K-k), \eeq
where $K$ is the radius of the phase plate in Fourier space. The goal of the phase plate is to shift the phase of the unscattered wave by $\pi/2$, bringing it into alignment with the scattered wave and thus giving maximum interference. The phase plate acts on the exit wave function to produce an image wave function according to
\beq \tilde \psi(\k) = \tilde \psi_\mathrm{exit}(\k)\tilde Z(\k).\eeq
For simplicity, we assume a perfect imaging system, and, unlike previous work \cite{DwyerPaganin2024}, here, for simplicity, we do not impose a high-spatial-frequency cutoff. Thus, in coordinate space, we obtain
\beq \psi(\rr) = (\psi_\mathrm{exit}\ast Z)(\rr),\eeq
where $\ast$ denotes convolution, $Z$ is the inverse Fourier transform of $\tilde Z$
\beq Z(\rr) = \delta(\rr) + (i-1)\tilde \theta(r),\eeq
and where, in a slight abuse of notation, $\tilde \theta$ is an Airy disc in coordinate space
\beq \tilde \theta(r) = \frac{K J_1(2\pi Kr)}{r}. \eeq
Using the expressions above, we obtain for the image wavefunction
\beq \psi(\rr) = \psi_0(\rr) + iV(\rr)\psi_0(\rr) +(i-1)\psi_1(\rr) - (i+1) \psi_2(\rr),\eeq
where 
\beq \psi_1(\rr) = (\psi_0\ast\tilde\theta)(\rr),\qquad \psi_2(\rr) = (V\psi_0\ast \tilde\theta)(\rr).\eeq
Consistent with the assumption of weak scattering, when calculating the image intensity $I(\rr) = |\psi(\rr)|^2$, we retain terms up to first order in $V$. Noting that $\psi_0$, $\psi_1$, $\psi_2$ and $V$ are real functions, we obtain, after some algebra,
\beq I(\rr) = |\psi_1(\rr)|^2 + 2\psi_0(\rr)\psi_1(\rr) V(\rr) + |\psi_0(\rr) - \psi_1(\rr)|^2   - 2\psi_0(\rr)\psi_2(\rr). \eeq

We are interested in the quantum and classical Fisher information \cite{Liu-etal2020} regarding a set of parameters $\lambda_1, \lambda_2,\dots$ which characterize the sample. The quantum Fisher information refers to the ``potential" information that is contained in the exit wave. In the present work, we do not evaluate the quantum Fisher information explicitly. The classical Fisher information refers to the ``actual" information contained in the recorded image. Using $\mu$ or $\nu$ to label the parameters, an element of the classical Fisher information can be written in the form
\beq F_{\mu\nu} = \frac{N}{\pi R^2} \int \d^2\rr\, \frac{I_\mu(\rr)I_\nu(\rr)}{I(\rr)}, \eeq
where $N$ is the number of electrons detected, $I_\mu$ denotes the partial derivative of the intensity with respect to parameter $\lambda_\mu$ (similarly for $I_\nu$), and we have divided by the area $\pi R^2$ of the incident beam to compensate for our (lack of) normalization. As we are interested in sample parameters, only the sample potential depends on the parameters. We obtain
\beq I_\mu(\rr) = 2\psi_0(\rr) \left(\psi_1(\rr)V_\mu(\rr) - \psi_{2,\mu}(\rr)\right).\eeq
For the Fisher information matrix element, we find, to leading order in $V$,
\beq\label{eq:Fisher finite beam} F_{\mu\nu} = \frac{4N}{\pi R^2} \int \d^2\rr\, \frac{ |\psi_0|^2(V_\mu\psi_1 - \psi_{2,\mu})(V_\nu\psi_1 - \psi_{2,\nu})}{ |\psi_1|^2 + |\psi_0-\psi_1|^2}. \eeq

Conventional Zernike imaging, which uses an (effectively) infinitely-wide incident plane wave and an infinitely-small phase plate, corresponds to the impulse limit $R\rightarrow\infty$ and $K\rightarrow0$ such that $KR\gg 1$ is fixed. In this limit, we obtain $\psi_1 \rightarrow \psi_0\rightarrow 1$ and $\psi_2 \rightarrow \av{V}$, giving
\beq I(\rr) = 1 + 2(V(\rr) - \av{V}),\eeq
which is the ideal Zernike image intensity. For the Fisher information, the impulse limit implies $\psi_{2,\mu}\rightarrow0$, and from Eq.~\eqref{eq:Fisher finite beam} we obtain \cite{Dwyer2023}
\beq F_{\mu\nu} = 4N V_\mu V_\nu . \eeq
This is the quantum limit, in which the classical Fisher information equals the quantum Fisher information, i.e., the maximum allowed by quantum mechanics.

For the case of a finite incident wave, we performed numerical calculations of the image intensity and Fisher information for a range of beam radii $R$ and phase plate cutoffs $K$, these being the two main parameters which characterize the departure from the conventional case.

\subsection{Measurement of atomic positions}

Here we choose the parameters $\set{\lambda_\mu}$ to be the positions of the atoms. This parameter type corresponds to features of the potential which are highly localized with respect to the width of the incident beam, and are therefore delocalized in Fourier space. We write the potential as a sum of atomic potentials in the form 
\beq V(\rr) = \sum_j V_j(\rr-\rr_j).\eeq
In fact, we need only consider a single, representative atom, and for its potential, we adopt a Gaussian form for simplicity
\beq V_j(\rr) =  A_j e^{-((x-x_j)^2 + (y-y_j)^2)/2\sigma^2}.\eeq
The parameters of the Gaussian are listed in Table~\ref{tab:atomic pos parameters}. The required derivatives are
\beq \frac{\del V}{\del x_j} = -V_{j,x}(\rr-\rr_j), \qquad \frac{\del V}{\del y_j} = -V_{j,y}(\rr-\rr_j) .\eeq

In the impulse limit, we obtain \cite{Dwyer2023}
\beq\label{eq:quantum limit atomic position} F_{\mu\nu} = \frac{4N \delta_{\mu\nu}}{\pi R^2}  \int\d^2\rr\, V^2_{j,x}(\rr-\rr_j), \qquad \lambda_\mu = x_j, \eeq
which is the quantum limit of Fisher information on $x_j$ for an incident electron fluence $N/\pi R^2$. An analogous result is obtained for $y_j$.

Table~\ref{tab:atomic pos} contains our results for the Fisher information from finite beams, which were calculated using Eq.~\eqref{eq:Fisher finite beam}. For each pair $R$ and $K$, we quote the Fisher information as a fraction of the quantum limit, as given by Eq.~\eqref{eq:quantum limit atomic position}. Selected calculated images and corresponding line trace plots are shown in Figs.~\ref{fig:atomic pos R 128}--\ref{fig:atomic pos R 512}. Each plot also shows the line trace of the difference image, which is obtained by subtracting the image with no sample. For each value of $R$, we observe a range of $K$ values where the Fisher information is $>95$\% of the quantum limit. Those are the cases where, in Figs.~\ref{fig:atomic pos R 128}--\ref{fig:atomic pos R 512}, the atom is seen with excellent contrast, that is, contrast which approaches the ideal case. For other cases, notably when $K$ is too small or too large, the contrast is diminished or lost, and so too is the Fisher information on the atom's position. For $K$ too small, the phase plate does not sufficiently contain the unscattered wave in order to rotate its phase by $\pi/2$. For $K$ too large, the phase plate contains too much of the scattered wave. Later we state the conditions on $R$ and $K$ which are needed for high Fisher information.

\subsection{Measurement of Fourier modulus and phase}

Here we choose the parameters $\set{\lambda_\mu}$ to be the moduli and phases of the potential Fourier coefficients. This parameter type corresponds to a feature of the potential which extends across the entire incident beam, and is therefore localized in Fourier space. We write the potential in the form 
\beq V(\rr) = \av{V} +  \sum_{\g>\0} 2|V_\g|\cos(2\pi \g\cdot\rr + \phi_\g),\eeq
where the sum is restricted to the half space (denoted $\g>\0$) on account of the relationship $V_\g = \bar{V}_{-\g}$. In fact, we need only consider a single, representative Fourier coefficient, whose parameters are listed in Table~\ref{tab:Fourier coeff parameters}. To evaluate the Fisher information, we will need the derivatives
\beq \frac{\del V}{\del|V_\g|} = 2\cos(2\pi \g\cdot\rr + \phi_\g), \qquad \frac{\del V}{\del\phi_\g} = -2 |V_\g| \sin(2\pi \g\cdot\rr + \phi_\g).\eeq
In the impulse limit, we obtain \cite{DwyerPaganin2024}
\beq F_{\mu\nu} = 8N \delta_{\mu\nu}\begin{cases} 1, & \text{$\lambda_\mu = |V_\g|$},\\
|V_\g|^2, & \text{$\lambda_\mu = \phi_\g$}, \end{cases} \eeq
which is the quantum limit. As the spatial frequency $|\g|$ increases, the Fisher information on $|V_\g|$ remains constant, whereas the Fisher information on $\phi_\g$ falls off according to $|V_\g|^2$.

Tables~\ref{tab:Fourier modulus} and \ref{tab:Fourier phase} contain our results for the Fisher information on the Fourier modulus and phase from finite beams. Selected calculated images and corresponding line trace plots are shown in Figs.~\ref{fig:Fourier coeff R 128}--\ref{fig:Fourier coeff R 512}. Similar to the case of atomic positions, for each value of $R$, we observe a range of $K$ values where the image shows excellent contrast and the Fisher information is $>95$\% of the quantum limit.

\section{Zernike operator}
\label{sec:Zernike operator}

In principle, it is possible to achieve the Zernike phase condition for any incident beam. This is easiest to formulate using abstract Dirac notation. For an arbitrary incident state $\ket{\psi_0}$, the exit state in the weak phase object approximation is
\beq \ket{\psi_\mathrm{exit}} = (\1 + i\V)\ket{\psi_0}. \eeq
We define the corresponding Zernike operator as
\beq\label{eq:Zernike operator} \Z = \1 + (i-1)\ketbra{\psi_0}{\psi_0}, \eeq
where $\ketbra{\psi_0}{\psi_0}$ is the projector onto the incident state. The Zernike operator $\op{Z}$ is unitary, as it should be. Acting on $\ket{\psi_\mathrm{exit}}$ with $\Z$, we get (omitting an unimportant overall factor $i$)
\beq \ket{\psi} = (1 + \V + (i-1)\av{V})\ket{\psi_0}. \eeq
From this, we obtain the ideal Zernike result for the intensity in coordinate space
\beq I(\rr) = |\braket{\rr}{\psi}|^2 = |\braket{\rr}{\psi_0}|^2 (1 + 2(\matel{\rr}{\V}{\rr}-\av{V}),\eeq
where $\av{V} = \matel{\psi_0}{\V}{\psi_0}$. Note the necessity of detecting the electrons in coordinate space where the potential is \emph{real} and \emph{diagonal}. If the detection is performed in some other plane, then the same effect cannot be achieved.

The main challenge in implementing the Zernike operator in Eq.~\eqref{eq:Zernike operator} lies in finding a physical implementation of the projector $\ketbra{\psi_0}{\psi_0}$. The projector must admit $\ket{\psi_0}$ and simultaneously block components orthogonal to $\ket{\psi_0}$. A physical implementation in terms of optical elements in coordinate space, Fourier space, or indeed some other space, is in most cases not obvious. An exception is conventional Zernike imaging, where the incident state is a plane wave $\ket{\psi_0} = \ket{\k_0}$, and the projector is diagonal in Fourier space
\beq \matel{\k}{\Z}{\k'} = \tilde Z(\k) \delta(\k-\k') = \left[1 + (i-1)\delta(\k-\k_0) \right] \delta(\k-\k'), \eeq
which represents a $\pi/2$ phase plate at $\k_0$.

\section{Zernike speckle imaging}
\label{sec:Zernike speckle imaging}

In this section, we consider an approximate implementation for a case where the incident wave differs substantially from a plane wave, namely, the case of a delocalized speckle wave. In previous work \cite{DwyerPaganin2024}, we showed that such an incident wave optimizes four-dimensional scanning transmission electron microscopy (4D STEM), in the sense that it can then deliver about half of the quantum limit across a large range of spatial frequencies. Recall that 4D STEM employs a pixellated detector to capture the scattering distribution in Fourier space \cite{Ophus2019}. Here, we consider using the same incident wave, but employ a pixellated detector in coordinate space.

\begin{figure} 
\includegraphics[width=0.3\columnwidth]{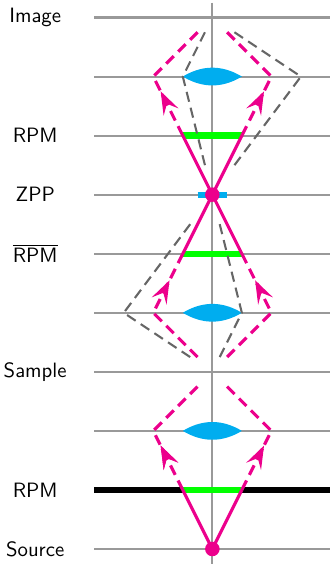}
\caption{\label{fig:Zernike speckle setup} Zernike speckle imaging setup using two identical random phase masks (RPMs), a conjugately-equivalent random phase mask (\lineovertext{RPM}), and a Zernike phase plate (ZPP).}
\end{figure}

We refer to Fig.~\ref{fig:Zernike speckle setup}, which shows a sketch of a Zernike speckle imaging setup based on two identical random phase masks (RPMs), a conjugately-equivalent random phase mask (\lineovertext{RPM}), and a Zernike phase plate (ZPP). The aperture is described by $A(\k) = \theta(K-k)$, where $K$ is the cutoff. The RPMs are described by
\beq R(\k) = 1 + (e^{i\phi(\k)}-1)\theta(K-k),\eeq
where $\phi(\k)$ is a random phase on $[0,2\pi)$. The \lineovertext{RPM} is described by $\bar R(\k)$ (if we were using successive forward, as opposed to forward and inverse, Fourier transforms to propagate through the optical system, we would describe \lineovertext{RPM} by $\bar R(-\k)$). The ZPP lies in coordinate space and is described by
\beq Z(\rr) = 1 + (i-1) \theta(R-r) , \eeq
where $R$ is the cutoff. 

The first RPM lies within the probe-forming aperture, and it gives rise to a speckled wave $\psi_0(\rr)$ incident on the sample. As before, we employ the weak phase-object approximation, which gives rise to a scattered wave $i(V(\rr)-V_\0)\psi_0(\rr)$ and an unscattered wave $(1+iV_\0)\psi_0(\rr)$, where $V_\0$ is the DC component, and we regard the associated term as part of the unscattered wave. 

The \lineovertext{RPM} ``despeckles" the unscattered wave, such that it forms a ``STEM probe" which is focused almost entirely within the ZPP where its phase is rotated by $\pi/2$. The final RPM then ``\emph{re}speckles" the unscattered wave, which interferes with the speckled scattered wave in the image plane. The scattered wave is displaced on \lineovertext{RPM}, thus it remains speckled in the plane of the ZPP, hence the majority of it passes outside the ZPP and experiences no phase change.

The entire optical setup is described by the following sequence of functions and Fourier transforms:
\beq\label{eq:Zernike speckle operator sequence} \F^{-1} R(\k) \F Z(\rr) \F^{-1} \bar{R}(\k) \F (1+iV(\rr)) \underbrace{\F^{-1}R(\k)A(\k)}_{\psi_0(\rr)}. \eeq
$\F$ and $\F^{-1}$ denote the Fourier transform and inverse Fourier transform defined previously. Despite its complicated appearance, Eq.~\eqref{eq:Zernike speckle operator sequence} readily simplifies. For the unscattered wave, the action of the ZPP is, approximately, to multiply by $i$ 
\beq \F^{-1} R \F \underbrace{Z}_{\approx i} \F^{-1} \bar{R} \F (1+iV_\0)\psi_0 \approx \underbrace{\F^{-1} R \F \F^{-1} \bar{R} \F}_{1}i(1+iV_\0) \psi_0 = i(1+iV_\0) \psi_0, \eeq
where $V_\0$ is the DC component. For the scattered wave, the action of the ZPP is, approximately, to multiply by $1$
\beq \F^{-1} R \F \underbrace{Z}_{\approx 1} \F^{-1} \bar{R} \F i(V-V_\0) \psi_0 \approx \underbrace{\F^{-1} R \F \F^{-1} \bar{R} \F}_{1} i(V-V_\0) \psi_0 = i(V-V_\0) \psi_0. \eeq
(The manipulations are valid for our form of $\psi_0$.) Thus, omitting the overall factor $i$, the image wave is, approximately, $(1+V(\rr)+(i-1)V_\0)\psi_0(\rr)$, and the image intensity is, approximately,
\beq I(\rr) \approx (1+2(V(\rr)-V_\0)) I_0(\rr).\eeq

\begin{figure} 
\includegraphics[width=0.3\columnwidth, trim=70 70 70 70, clip]{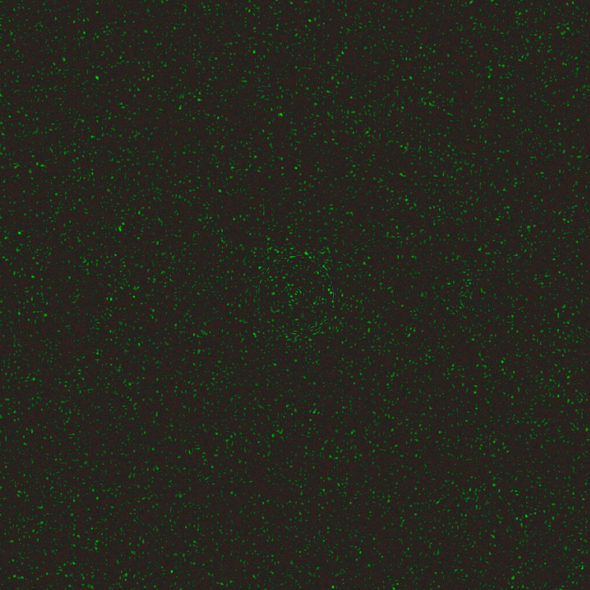}
\includegraphics[width=0.3\columnwidth, trim=70 70 70 70, clip]{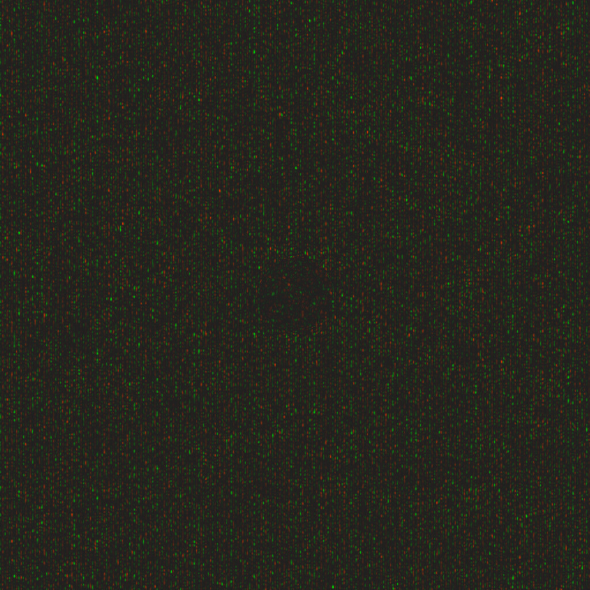}
\caption{\label{fig:Zernike speckle images} Zernike speckle image (left) and the difference image (right) obtained by subtracting the vacuum image from the speckle image. The difference image reveals the lattice fringes corresponding to $V_\g$. A convergence cutoff $K=128/2048$ and a ZPP radius $R=128$ were used.}
\end{figure}

For the speckle imaging setup described above, we calculated the Fisher information on the Fourier modulus and phase, using the parameters for the Fourier coefficient given previously, and a probe aperture cutoff $K=128/2048$. The results are given in Table~\ref{tab:speckle} and Fig.~\ref{fig:Zernike speckle images}.

We note that the speckle intensity $I_0(\rr)$ contains zeros which could mask localized features of $V(\rr)$. This could be mitigated by translating the incident wave across the sample, as in STEM, and recording the image for each translation. For example, writing $I(\rr,\rr_0)$ and $I_0(\rr,\rr_0)$ for the intensities corresponding to a translation $\rr_0$, the potential can be extracted following a procedure analogous to that outlined by Schiske (2002) \cite{Schiske2002}
\beq \frac{\sum_{\rr_0} \left(I(\rr,\rr_0)-I_0(\rr,\rr_0)\right)}{2 \sum_{\rr_0} I_0(\rr,\rr_0)} = V(\rr) - V_\0, \eeq
where the denominator is always positive, and thus the expression is well behaved.

\section{Discussion}
\label{sec:discussion}

We have shown that Zernike phase-contrast imaging using a finite beam and a finite phase plate in Fourier space can achieve $>95$\% of the quantum limit (with our stated idealizations). For the method to be effective, the majority of the unscattered wave must pass through the phase plate, which imposes the first condition $KR\gtrsim1$. Our results (Tables~\ref{tab:atomic pos}, \ref{tab:Fourier modulus}, \ref{tab:Fourier phase} and \ref{tab:speckle}) indicate that $RK \ge 1$ is effective (recall the first zero of the Airy disc Eq.~\eqref{eq:psi0k} occurs at $Rk = 0.61$).

On the other hand, the majority of the scattered wave must pass \emph{outside} the phase plate, and thus the second condition $K < K_s$, where $K_s$ is a spatial frequency of interest. Both conditions can be satisfied only if $R\ge1/K_s$, i.e., the beam must be sufficiently wide compared to the (inverse of the) spatial frequency of interest. Usually, TEMs can produce such a beam quite readily. Information on the very lowest spatial frequencies $\ll 1/R$ is inevitably lost.

\begin{figure}
\hspace{15mm} Ideal Case 
\includegraphics[width=0.24\columnwidth]{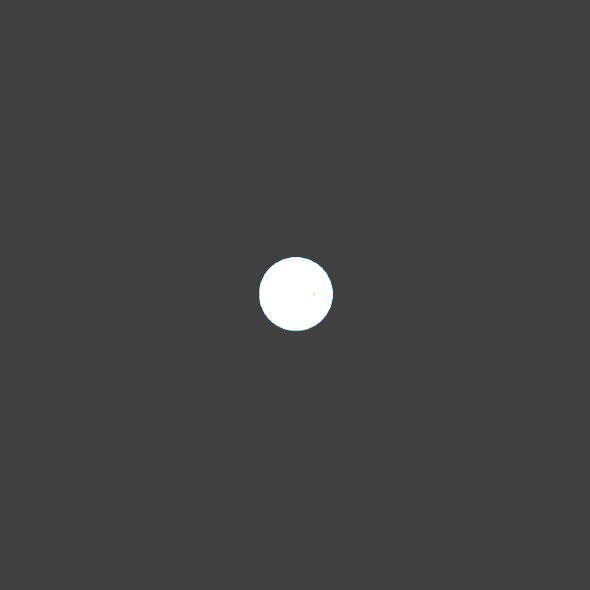}
\includegraphics[width=0.405\columnwidth, trim=30 28 12 22, clip]{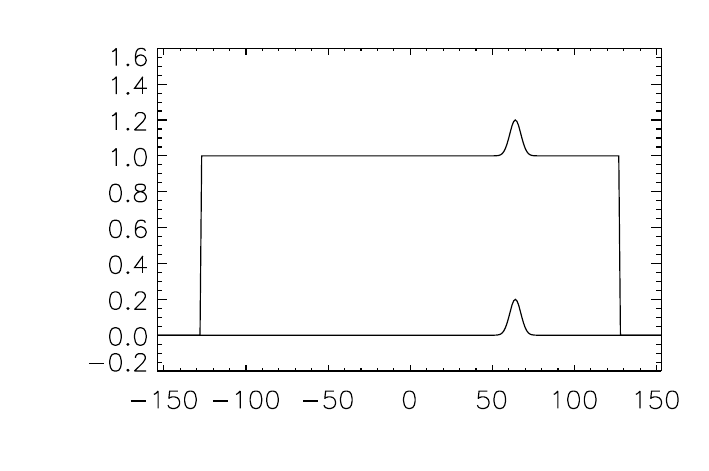}\\\vspace{4mm}
$K = \tfrac{4}{2048}$, $RK = 0.25$
\includegraphics[width=0.24\columnwidth]{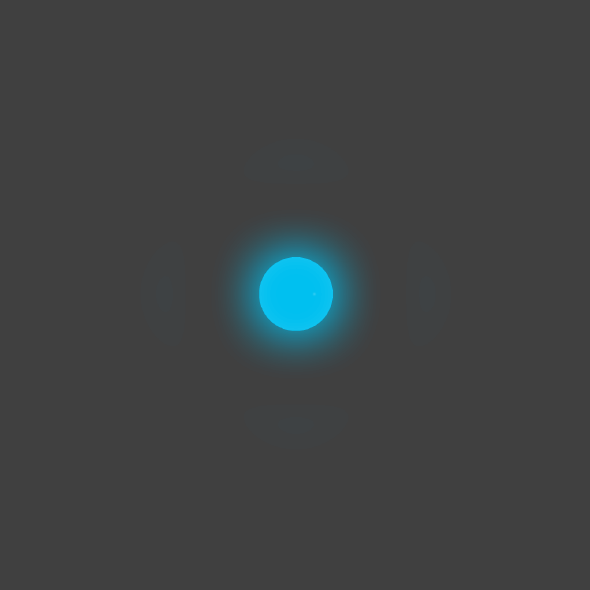}
\includegraphics[width=0.405\columnwidth, trim=30 28 12 22, clip]{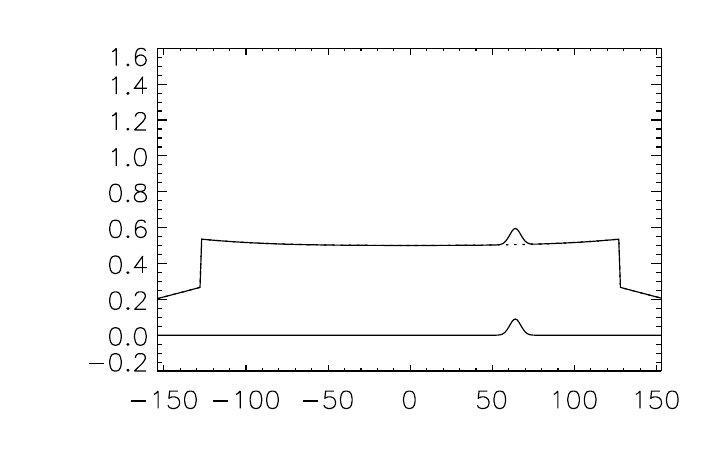}\\\vspace{4mm}
$K = \tfrac{16}{2048}$, $RK = 1.00$
\includegraphics[width=0.24\columnwidth]{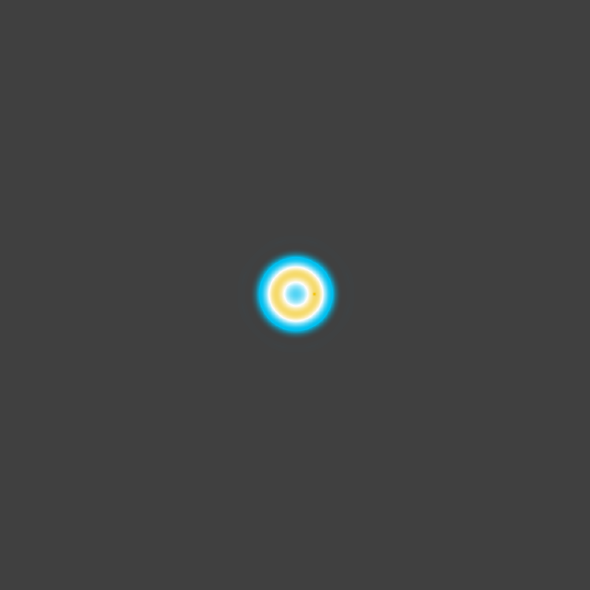}
\includegraphics[width=0.405\columnwidth, trim=30 28 12 22, clip]{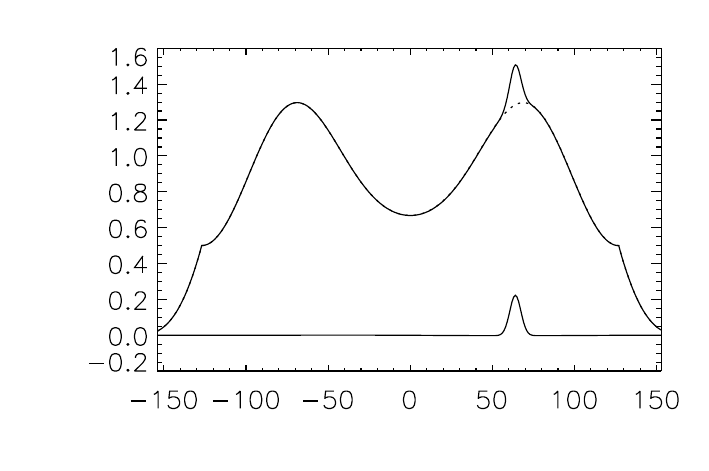}\\\vspace{4mm}
$K = \tfrac{64}{2048}$, $RK = 4.00$
\includegraphics[width=0.24\columnwidth]{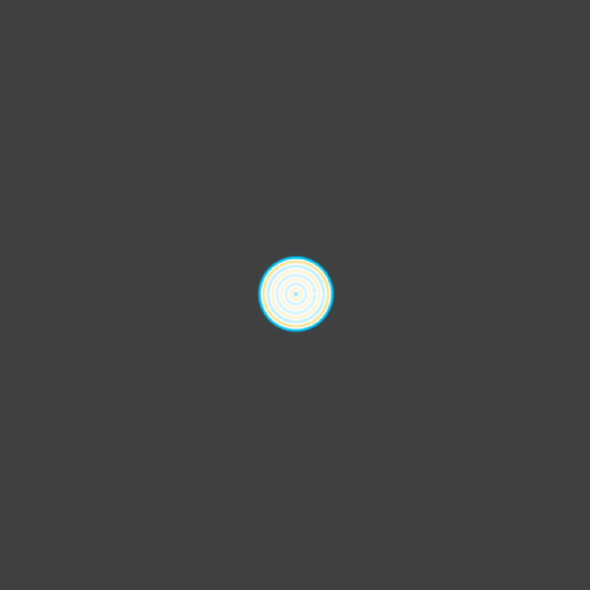}
\includegraphics[width=0.405\columnwidth, trim=30 28 12 22, clip]{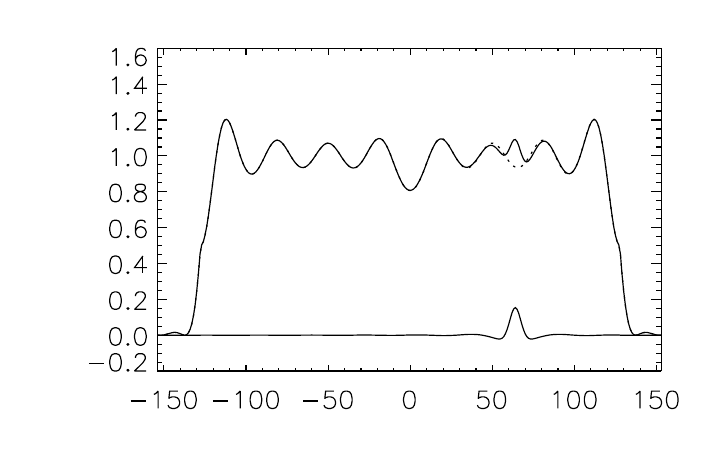}\\\vspace{4mm}
$K = \tfrac{256}{2048}$, $RK = 16.0$
\includegraphics[width=0.24\columnwidth]{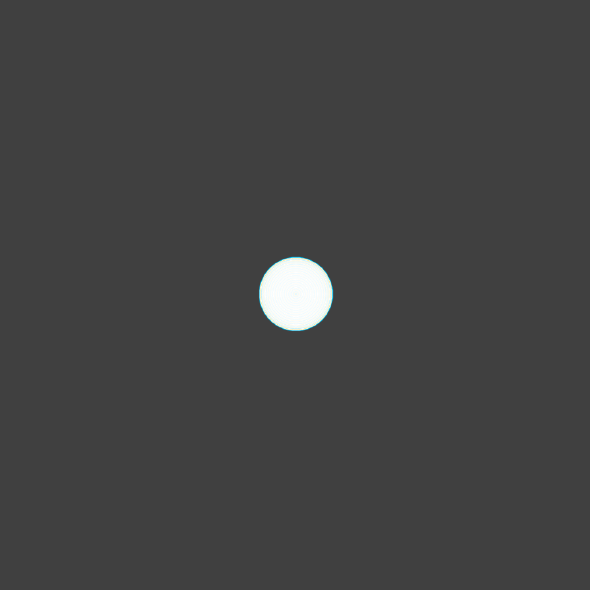}
\includegraphics[width=0.405\columnwidth, trim=30 28 12 22, clip]{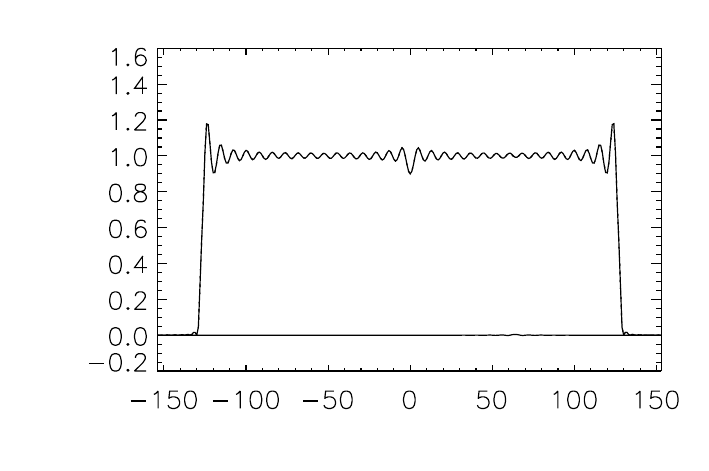}
\caption{\label{fig:atomic pos R 128} Zernike images for measurement of atomic position, using $R=128$. See Tables~\ref{tab:atomic pos parameters} and \ref{tab:atomic pos}.}
\end{figure}

\begin{figure}
\hspace{15mm} Ideal Case 
\includegraphics[width=0.24\columnwidth]{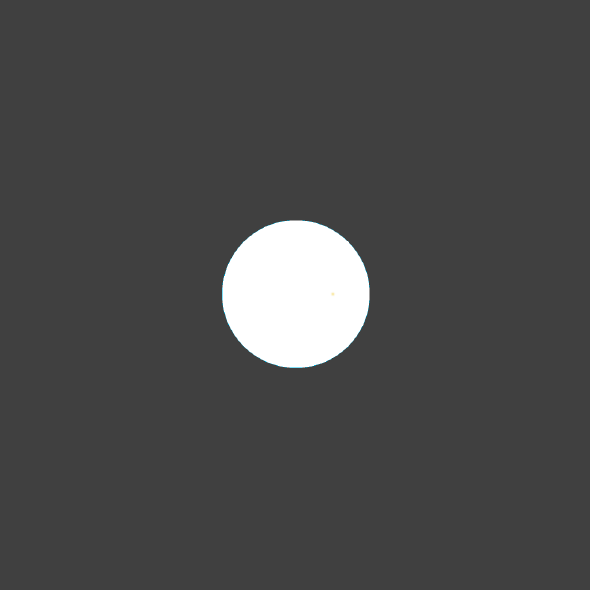}
\includegraphics[width=0.405\columnwidth, trim=30 28 12 22, clip]{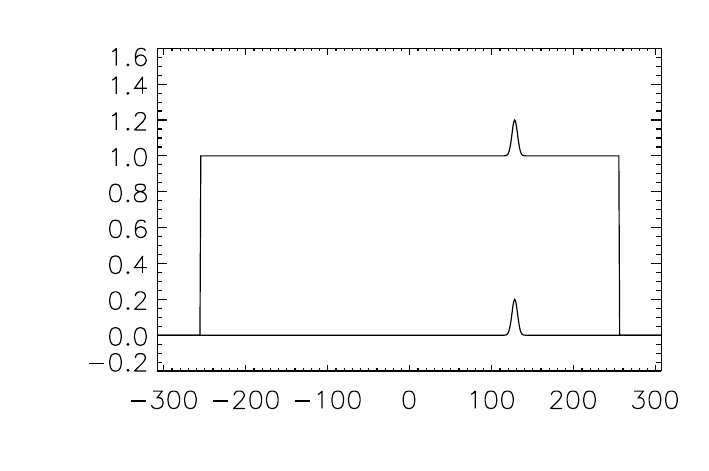}\\\vspace{4mm}
$K = \tfrac{4}{2048}$, $RK = 0.50$
\includegraphics[width=0.24\columnwidth]{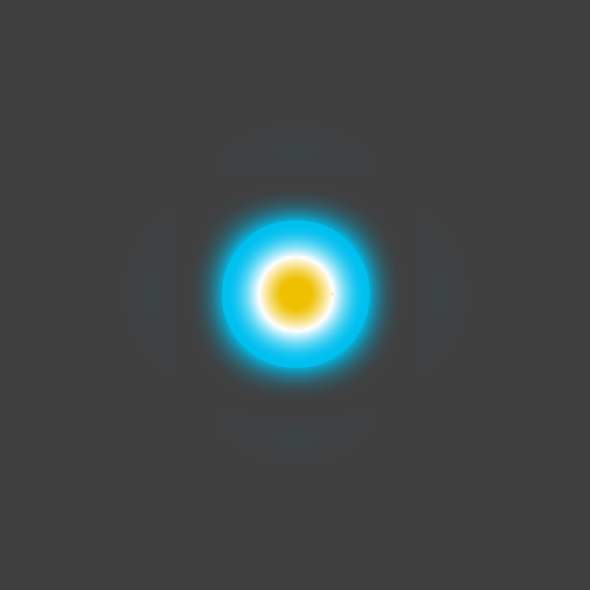}
\includegraphics[width=0.405\columnwidth, trim=30 28 12 22, clip]{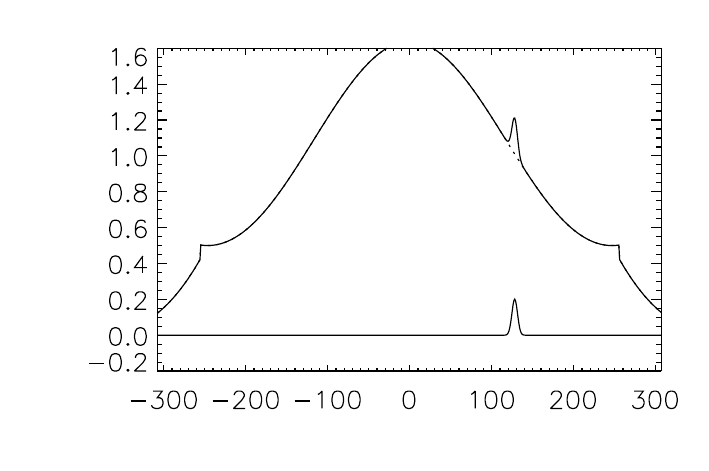}\\\vspace{4mm}
$K = \tfrac{16}{2048}$, $RK = 2.00$
\includegraphics[width=0.24\columnwidth]{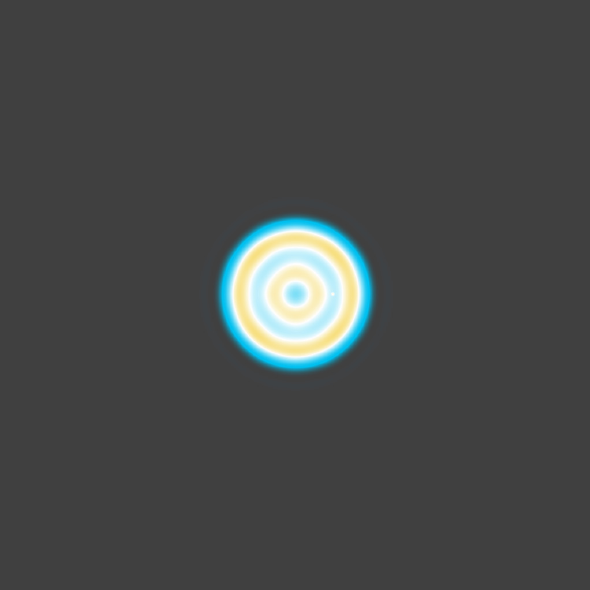}
\includegraphics[width=0.405\columnwidth, trim=30 28 12 22, clip]{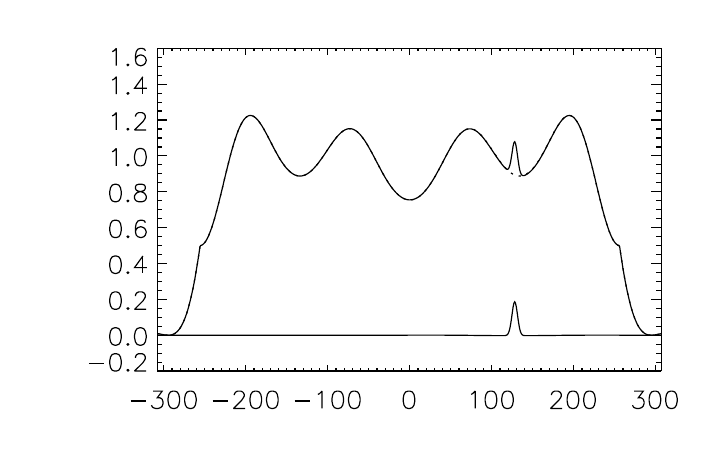}\\\vspace{4mm}
$K = \tfrac{64}{2048}$, $RK = 8.00$
\includegraphics[width=0.24\columnwidth]{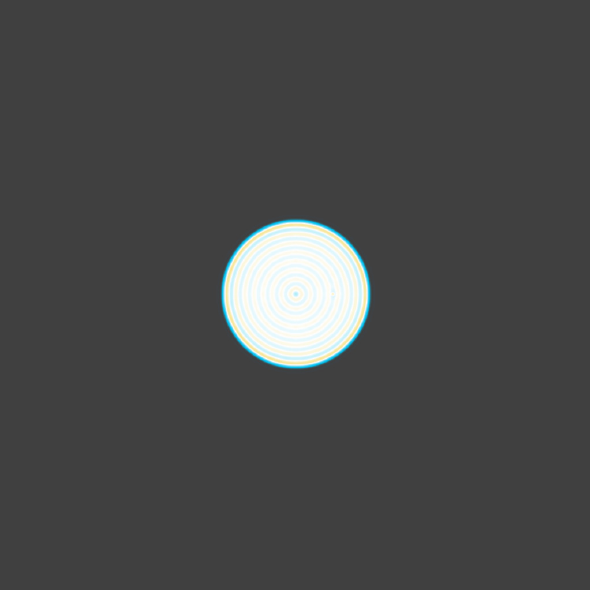}
\includegraphics[width=0.405\columnwidth, trim=30 28 12 22, clip]{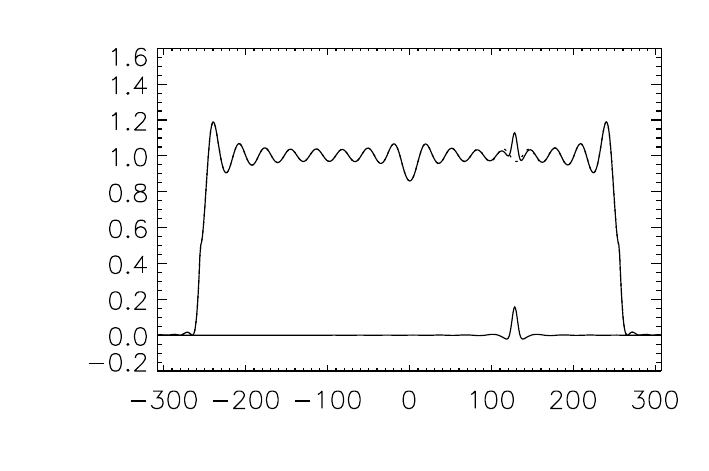}\\\vspace{4mm}
$K = \tfrac{256}{2048}$, $RK = 32.0$
\includegraphics[width=0.24\columnwidth]{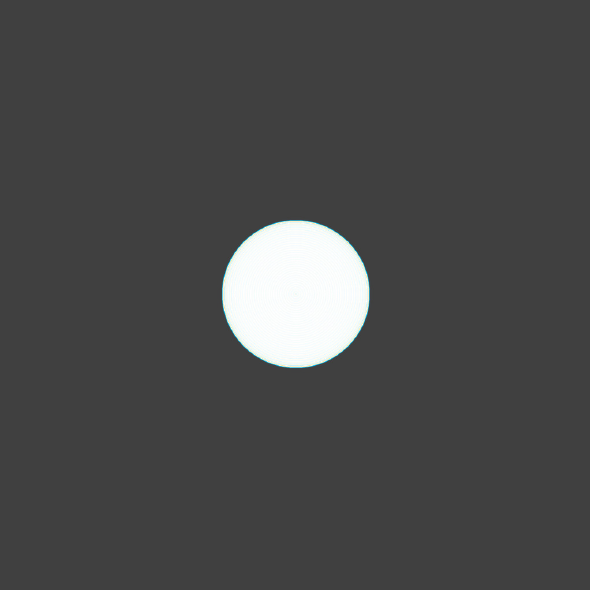}
\includegraphics[width=0.405\columnwidth, trim=30 28 12 22, clip]{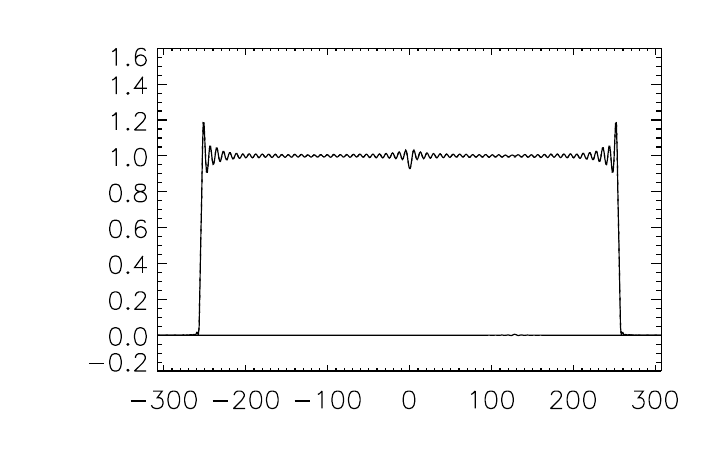}
\caption{\label{fig:atomic pos R 256} Zernike images for measurement of atomic position, using $R=256$. See Tables~\ref{tab:atomic pos parameters} and \ref{tab:atomic pos}.}
\end{figure}

\begin{figure}
\hspace{15mm} Ideal Case 
\includegraphics[width=0.24\columnwidth]{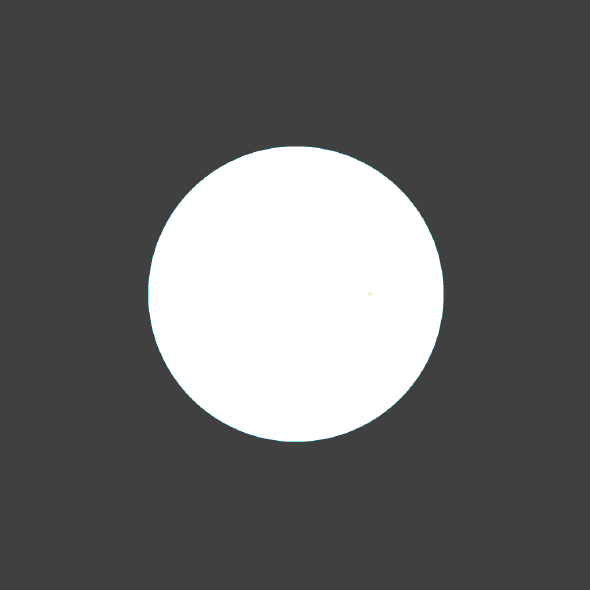}
\includegraphics[width=0.405\columnwidth, trim=30 28 12 22, clip]{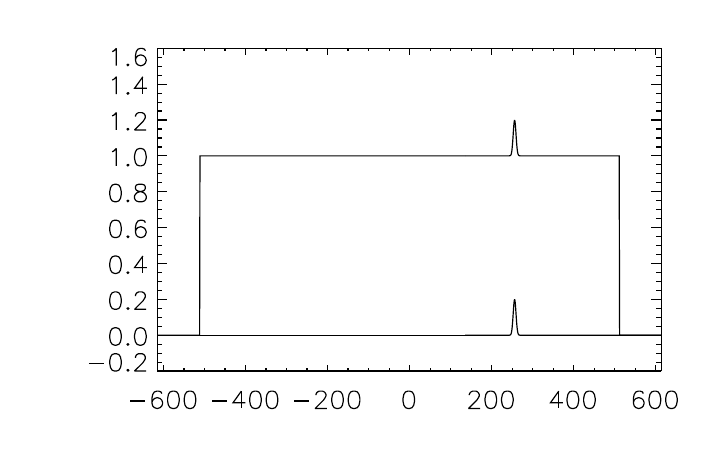}\\\vspace{4mm}
$K = \tfrac{4}{2048}$, $RK = 1.00$
\includegraphics[width=0.24\columnwidth]{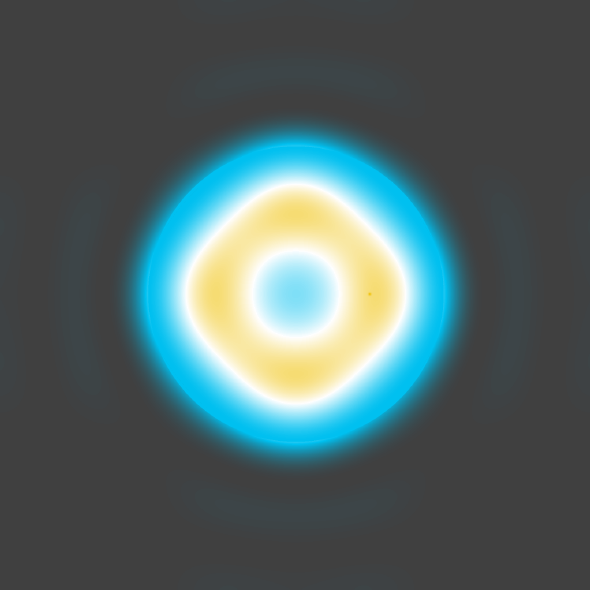}
\includegraphics[width=0.405\columnwidth, trim=30 28 12 22, clip]{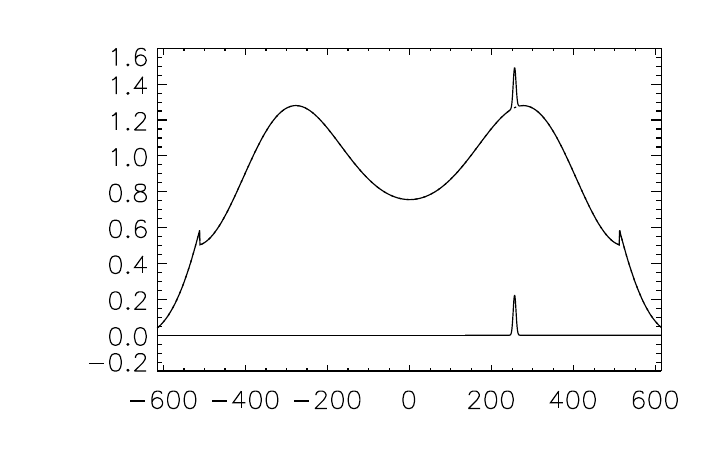}\\\vspace{4mm}
$K = \tfrac{16}{2048}$, $RK = 4.00$
\includegraphics[width=0.24\columnwidth]{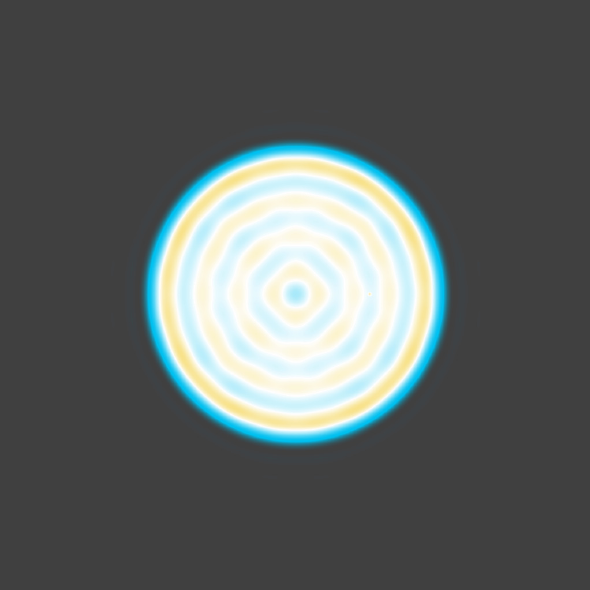}
\includegraphics[width=0.405\columnwidth, trim=30 28 12 22, clip]{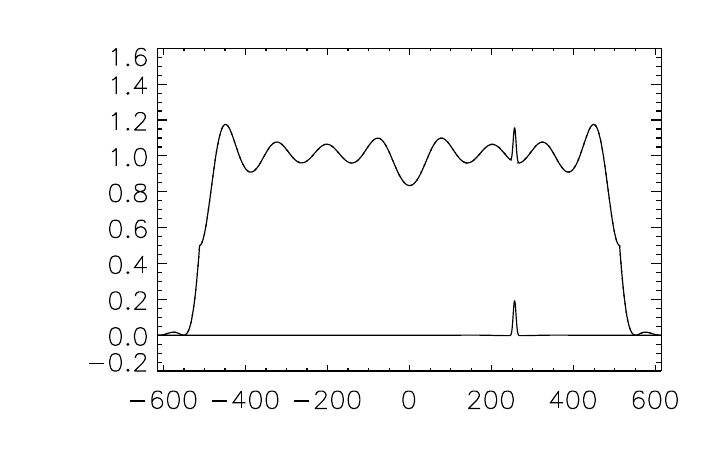}\\\vspace{4mm}
$K = \tfrac{64}{2048}$, $RK = 16.0$
\includegraphics[width=0.24\columnwidth]{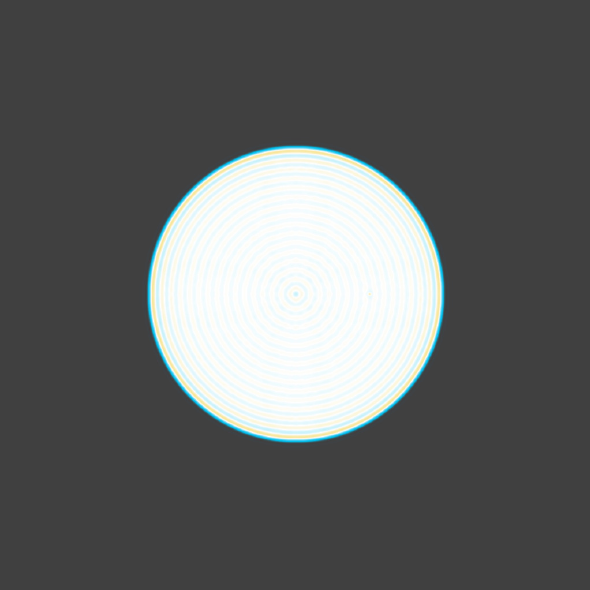}
\includegraphics[width=0.405\columnwidth, trim=30 28 12 22, clip]{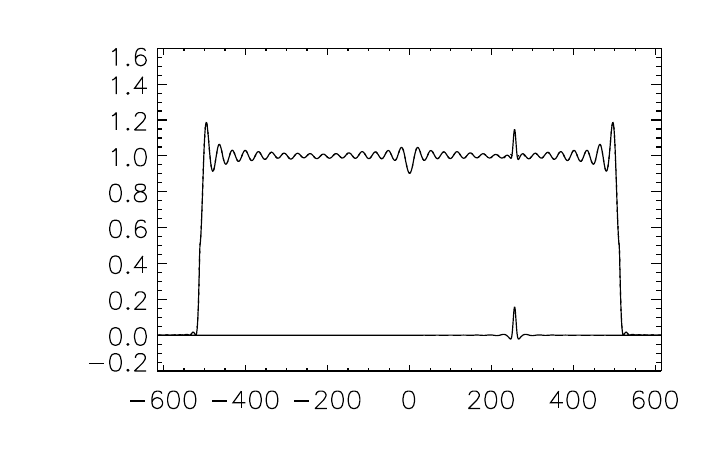}\\\vspace{4mm}
$K = \tfrac{256}{2048}$, $RK = 64.0$
\includegraphics[width=0.24\columnwidth]{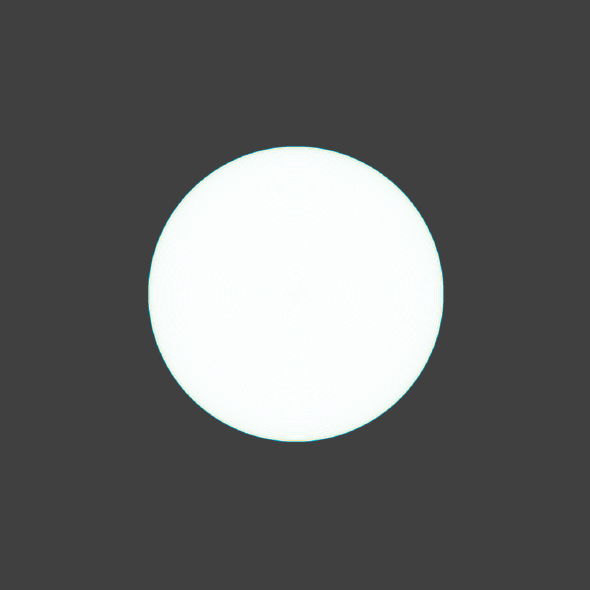}
\includegraphics[width=0.405\columnwidth, trim=30 28 12 22, clip]{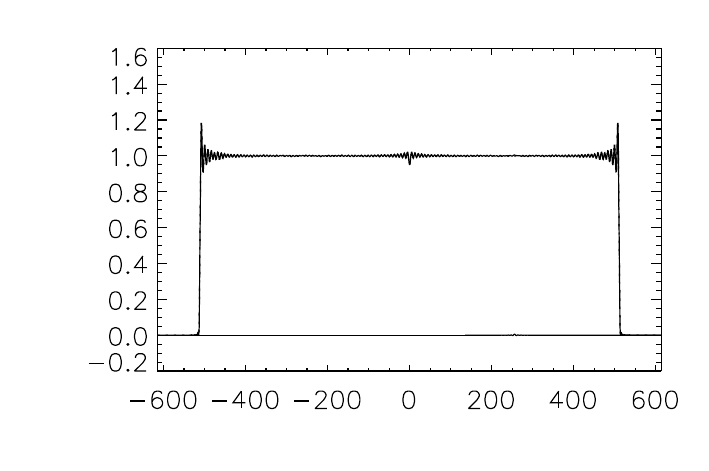}
\caption{\label{fig:atomic pos R 512} Zernike images for measurement of atomic position, using $R=512$. See Tables~\ref{tab:atomic pos parameters} and \ref{tab:atomic pos}.}
\end{figure}

\begin{figure}
\hspace{15mm} Ideal Case 
\includegraphics[width=0.24\columnwidth]{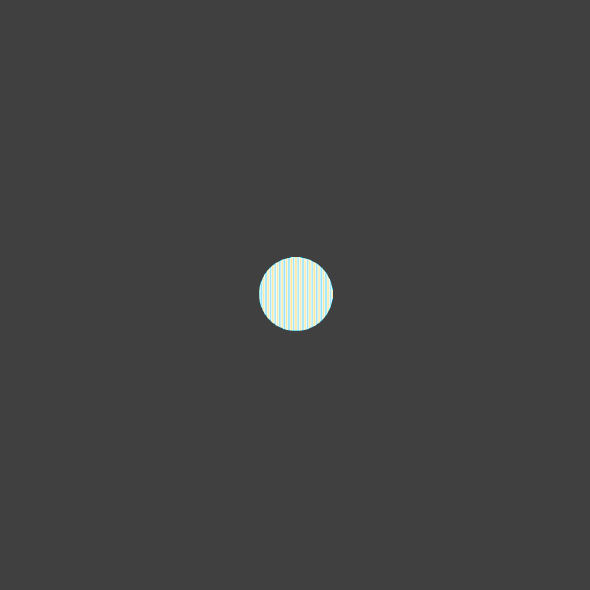}
\includegraphics[width=0.405\columnwidth, trim=30 28 12 22, clip]{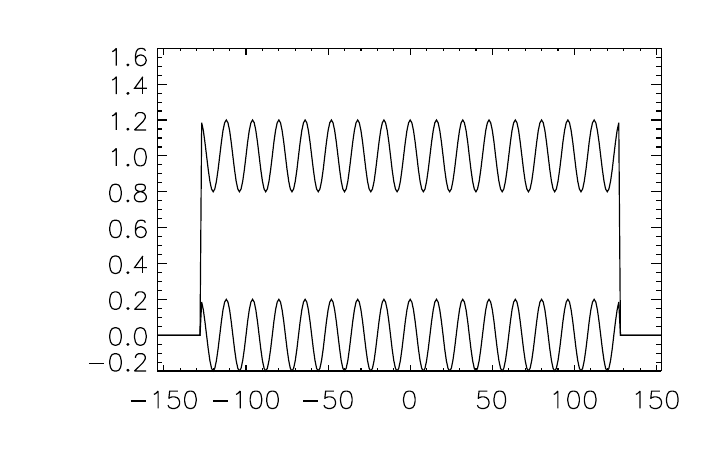}\\\vspace{4mm}
$K = \tfrac{4}{2048}$, $RK = 0.25$
\includegraphics[width=0.24\columnwidth]{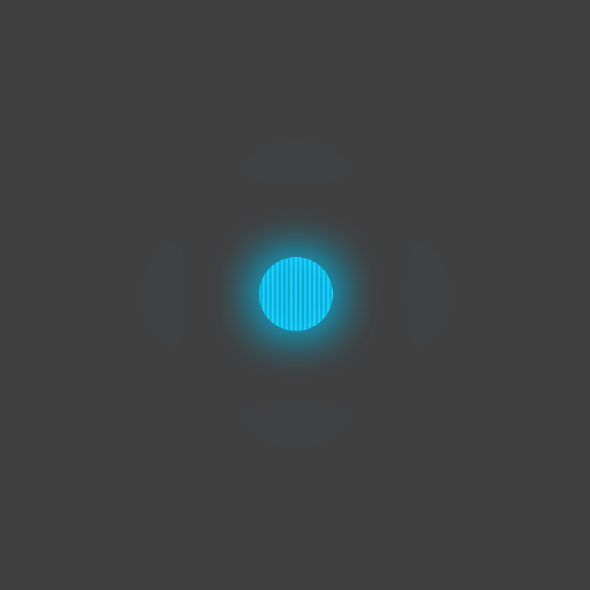}
\includegraphics[width=0.405\columnwidth, trim=30 28 12 22, clip]{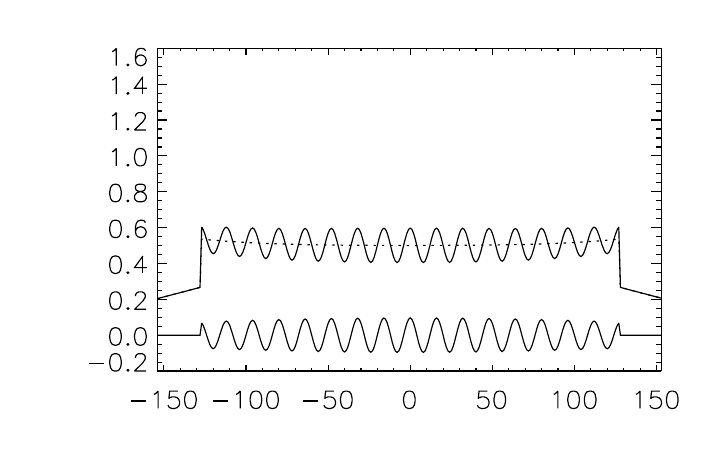}\\\vspace{4mm}
$K = \tfrac{16}{2048}$, $RK = 1.00$
\includegraphics[width=0.24\columnwidth]{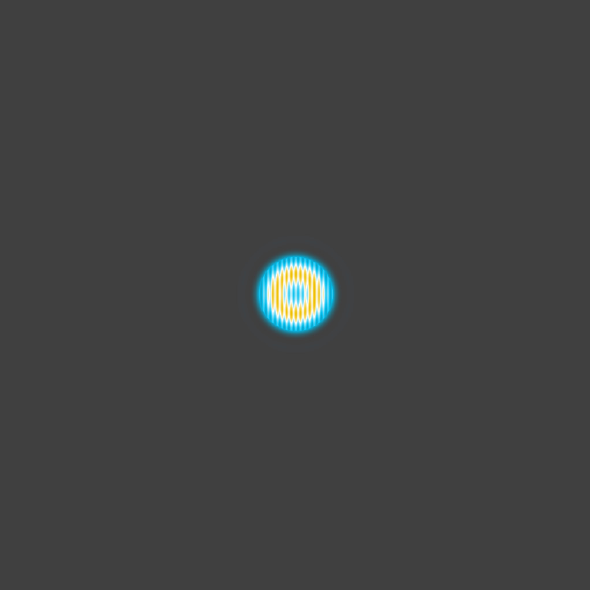}
\includegraphics[width=0.405\columnwidth, trim=30 28 12 22, clip]{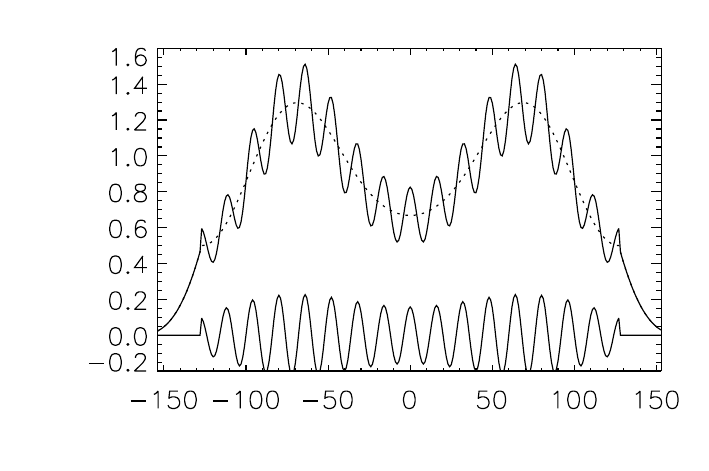}\\\vspace{4mm}
$K = \tfrac{64}{2048}$, $RK = 4.00$
\includegraphics[width=0.24\columnwidth]{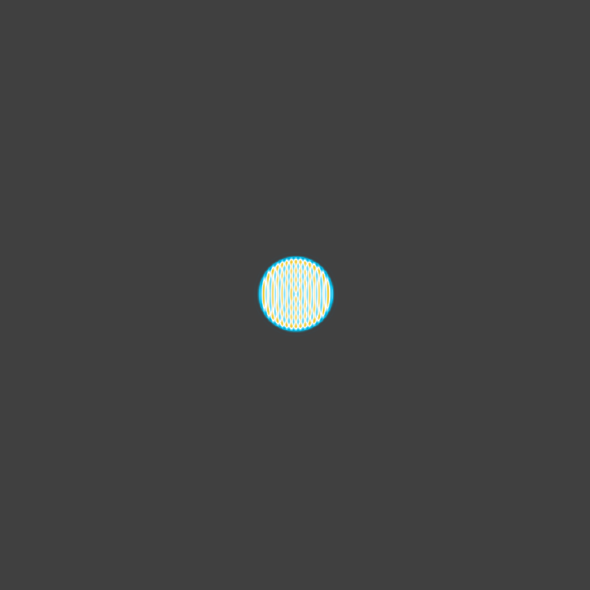}
\includegraphics[width=0.405\columnwidth, trim=30 28 12 22, clip]{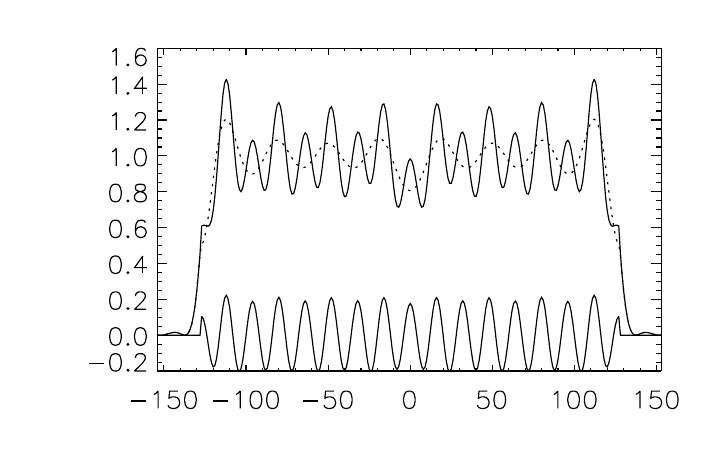}\\\vspace{4mm}
$K = \tfrac{256}{2048}$, $RK = 16.0$
\includegraphics[width=0.24\columnwidth]{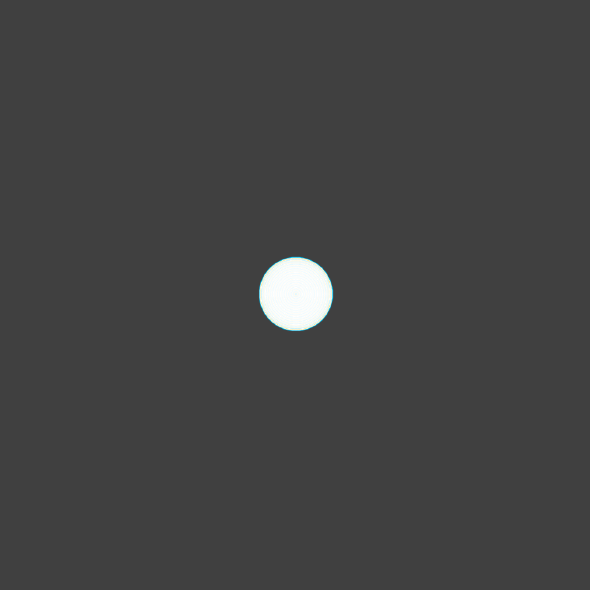}
\includegraphics[width=0.405\columnwidth, trim=30 28 12 22, clip]{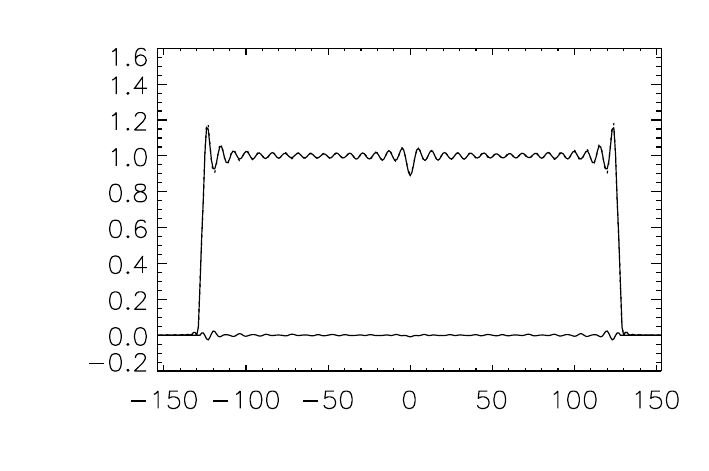}
\caption{\label{fig:Fourier coeff R 128} Zernike images for measurement of Fourier coefficient, using $R=128$. See Tables~\ref{tab:Fourier coeff parameters}, \ref{tab:Fourier modulus} and \ref{tab:Fourier phase}.}
\end{figure}

\begin{figure}
\hspace{15mm} Ideal Case 
\includegraphics[width=0.24\columnwidth]{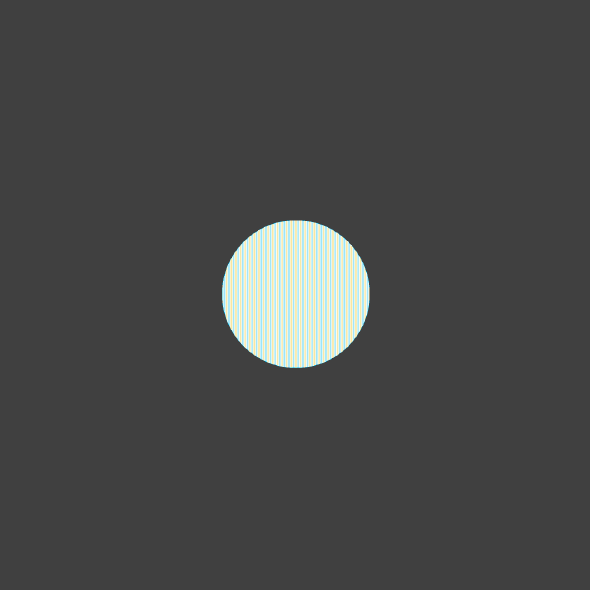}
\includegraphics[width=0.405\columnwidth, trim=30 28 12 22, clip]{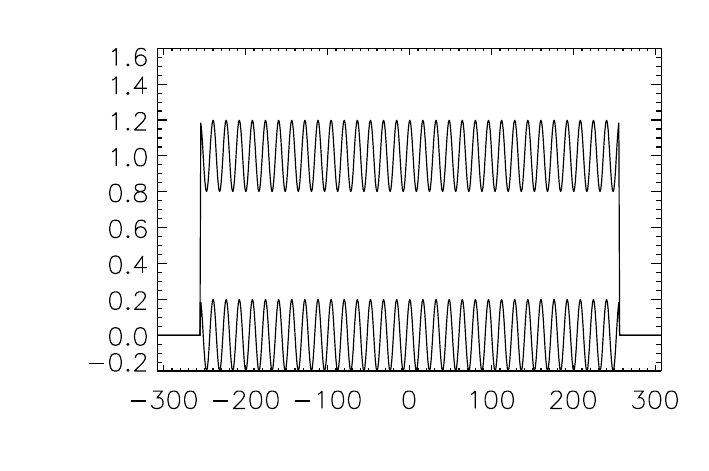}\\\vspace{4mm}
$K = \tfrac{4}{2048}$, $RK = 0.50$
\includegraphics[width=0.24\columnwidth]{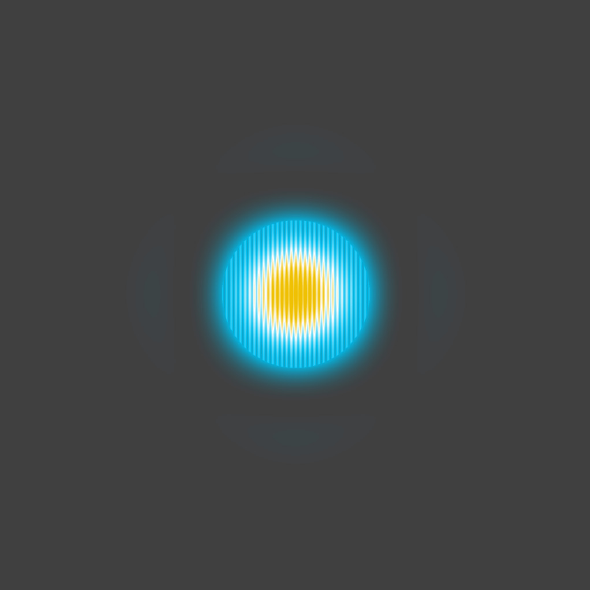}
\includegraphics[width=0.405\columnwidth, trim=30 28 12 22, clip]{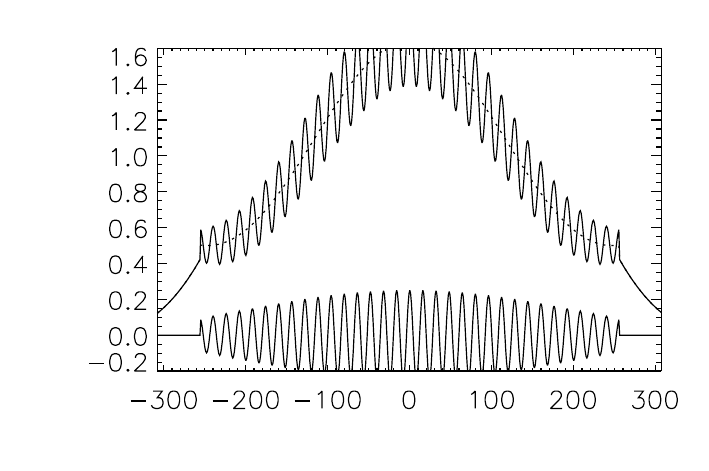}\\\vspace{4mm}
$K = \tfrac{16}{2048}$, $RK = 2.00$
\includegraphics[width=0.24\columnwidth]{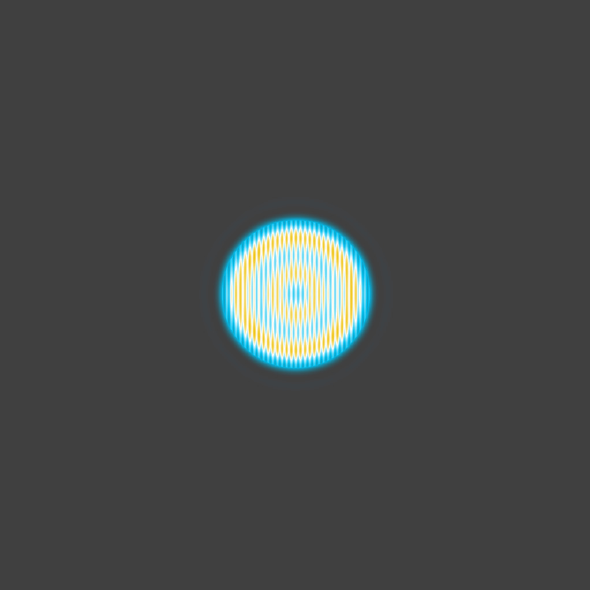}
\includegraphics[width=0.405\columnwidth, trim=30 28 12 22, clip]{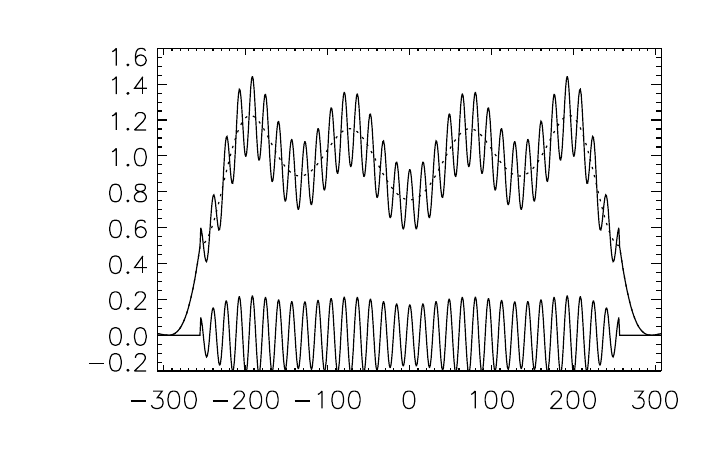}\\\vspace{4mm}
$K = \tfrac{64}{2048}$, $RK = 8.00$
\includegraphics[width=0.24\columnwidth]{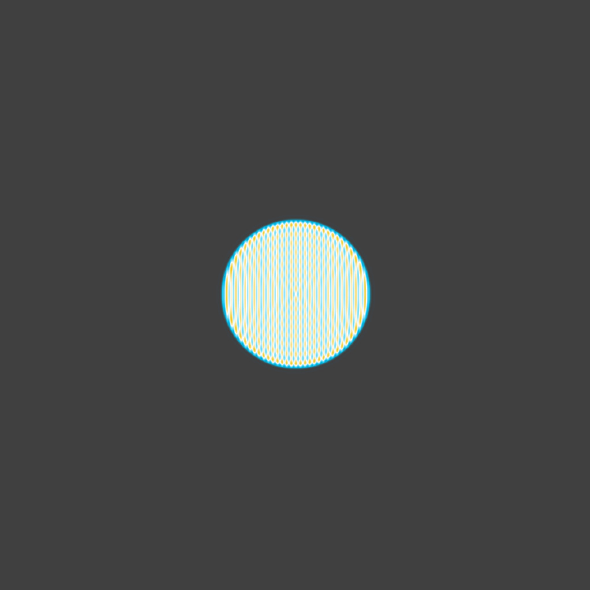}
\includegraphics[width=0.405\columnwidth, trim=30 28 12 22, clip]{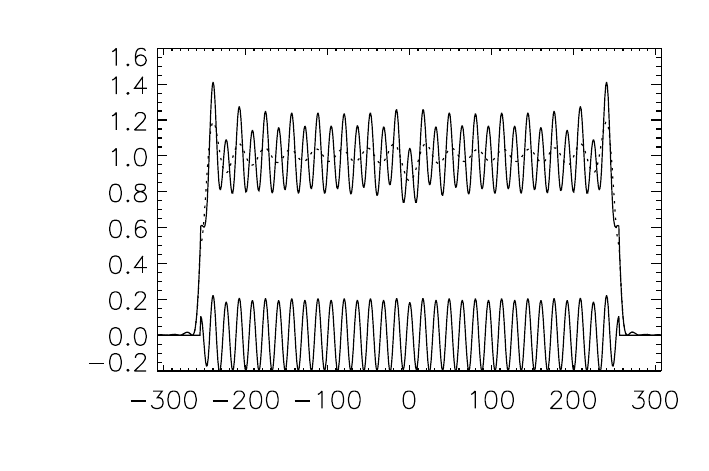}\\\vspace{4mm}
$K = \tfrac{256}{2048}$, $RK = 32.0$
\includegraphics[width=0.24\columnwidth]{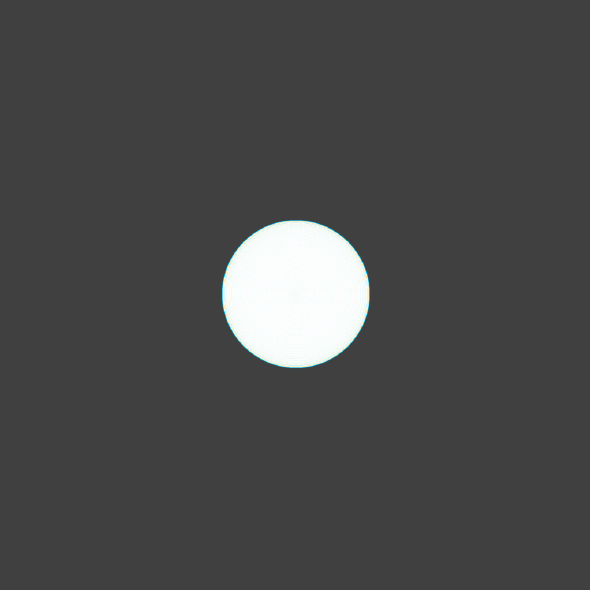}
\includegraphics[width=0.405\columnwidth, trim=30 28 12 22, clip]{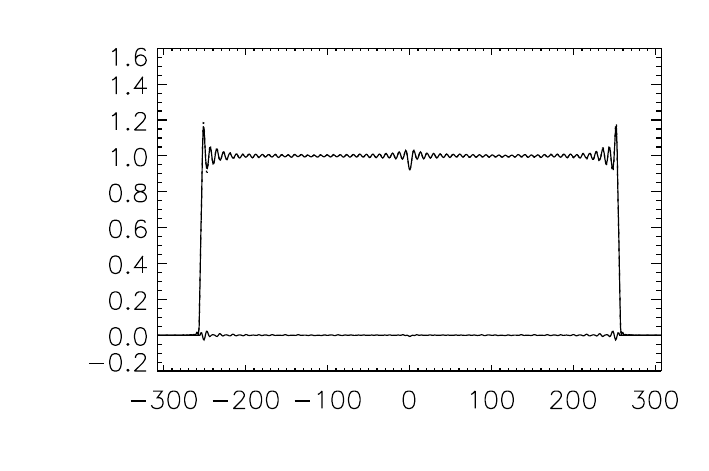}
\caption{\label{fig:Fourier coeff R 256} Zernike images for measurement of Fourier coefficient, using $R=256$. See Tables~\ref{tab:Fourier coeff parameters}, \ref{tab:Fourier modulus} and \ref{tab:Fourier phase}.}
\end{figure}

\begin{figure}
\hspace{15mm} Ideal Case 
\includegraphics[width=0.24\columnwidth]{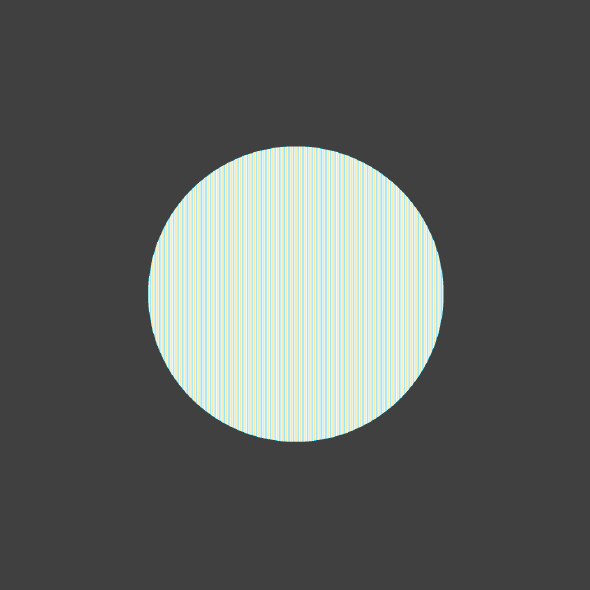}
\includegraphics[width=0.405\columnwidth, trim=30 28 12 22, clip]{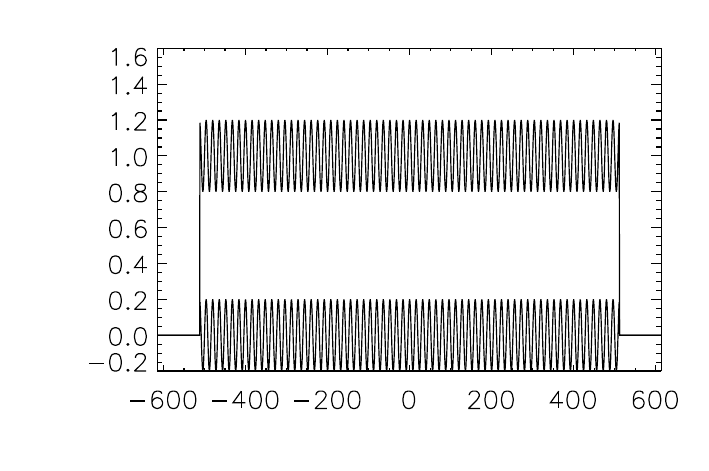}\\\vspace{4mm}
$K = \tfrac{4}{2048}$, $RK = 1.00$
\includegraphics[width=0.24\columnwidth]{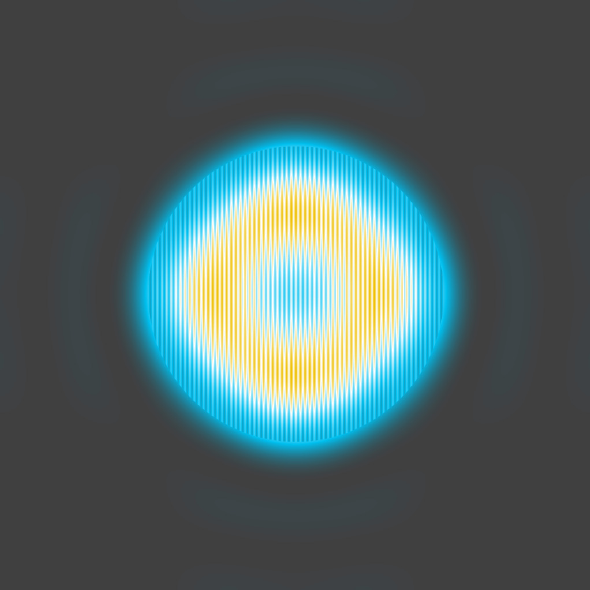}
\includegraphics[width=0.405\columnwidth, trim=30 28 12 22, clip]{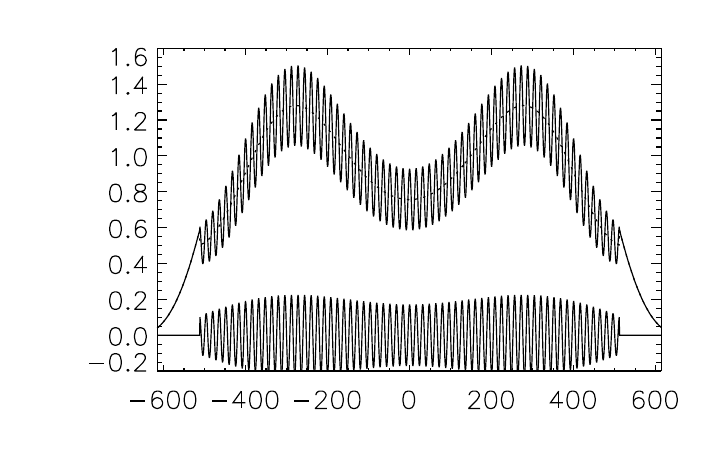}\\\vspace{4mm}
$K = \tfrac{16}{2048}$, $RK = 4.00$
\includegraphics[width=0.24\columnwidth]{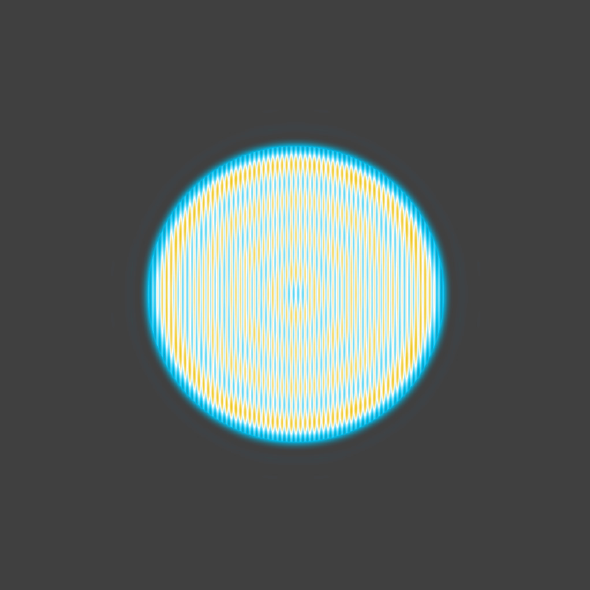}
\includegraphics[width=0.405\columnwidth, trim=30 28 12 22, clip]{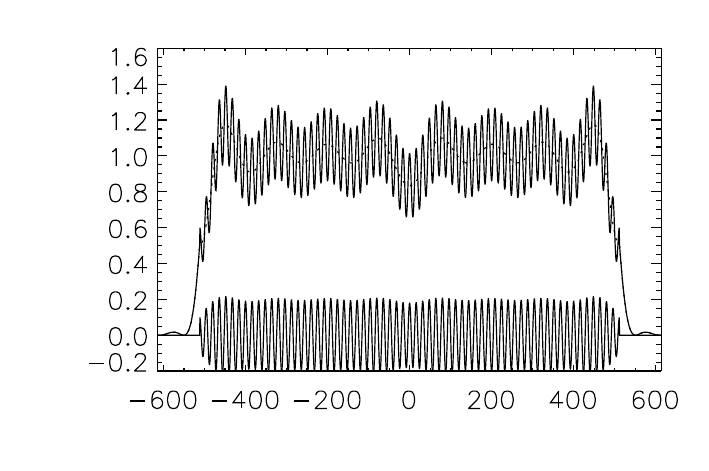}\\\vspace{4mm}
$K = \tfrac{64}{2048}$, $RK = 16.0$
\includegraphics[width=0.24\columnwidth]{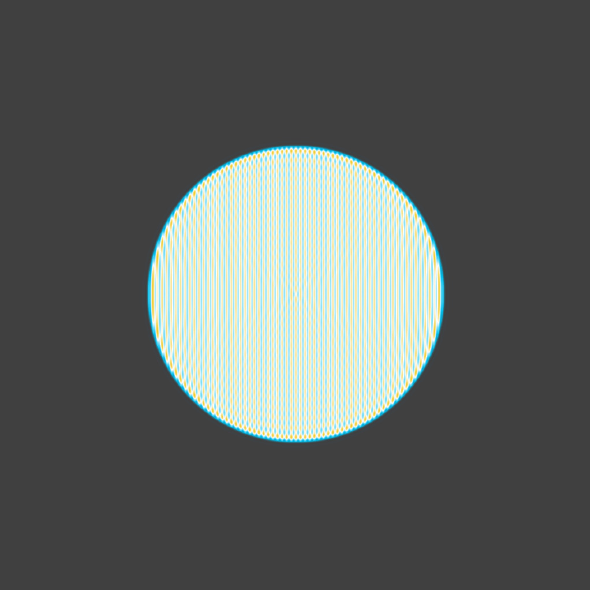}
\includegraphics[width=0.405\columnwidth, trim=30 28 12 22, clip]{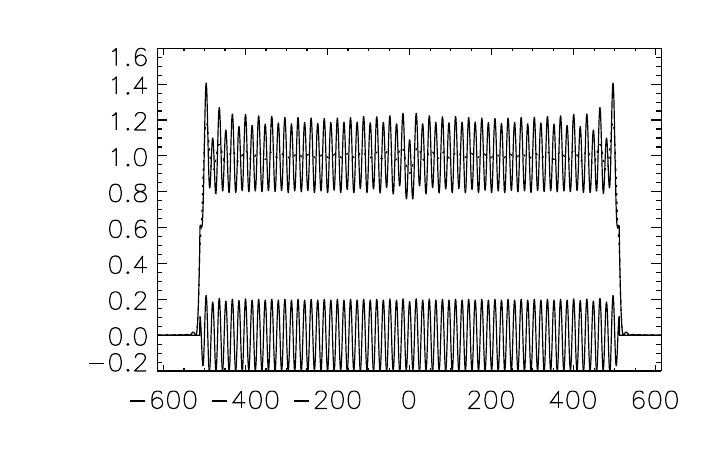}\\\vspace{4mm}
$K = \tfrac{256}{2048}$, $RK = 64.0$
\includegraphics[width=0.24\columnwidth]{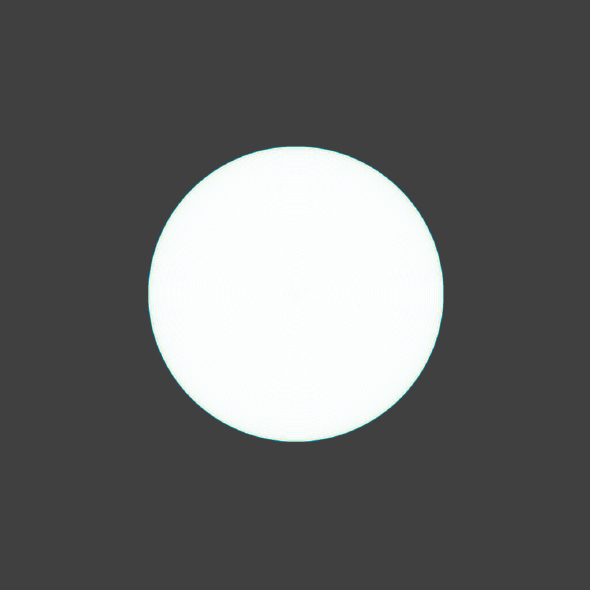}
\includegraphics[width=0.405\columnwidth, trim=30 28 12 22, clip]{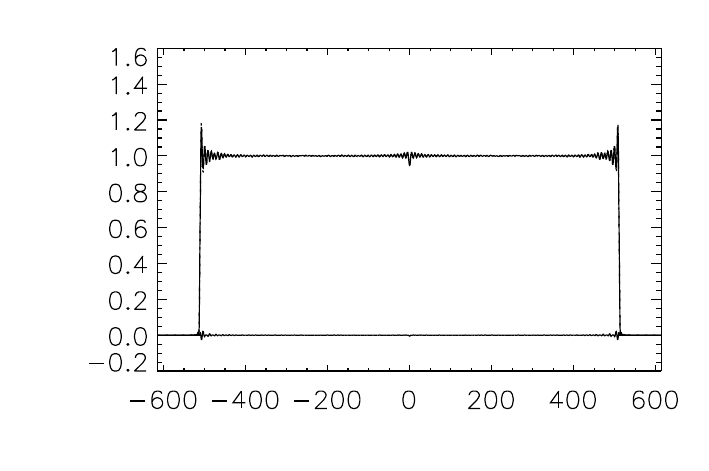}
\caption{\label{fig:Fourier coeff R 512} Zernike images for measurement of Fourier coefficient, using $R=512$. See Tables~\ref{tab:Fourier coeff parameters}, \ref{tab:Fourier modulus} and \ref{tab:Fourier phase}.}
\end{figure}

Achieving uniform illumination in the image plane requires $RK \ge 1$, and in that case it may well be that the spatial frequencies of interest are lost, such as occurs in Figs.~\ref{fig:atomic pos R 128}--\ref{fig:Fourier coeff R 512} for the largest values of $K$. However, uniform illumination should be viewed as a luxury rather than a necessity, as the Zernike method is effective regardless. We emphasize that the Fisher information provides a measure of information based (solely) on the sensitivity of the underlying probability distribution to the value of the parameter. It is not prejudiced by the appearance of the image. It has no regard as to whether humans can easily interpret the information or not.

We note that Vega Ib{\'a}{\~n}ez and Verbeeck (2024) \cite{VegaIbanezVerbeeck2025} reached rather different conclusions regarding the effectiveness of the Zernike method for a finite beam. However, those authors used $RK=1/2$, meaning that the phase plate was too small, and so it is unsurprising that they found suboptimal performance. As we have shown, we require $KR\gtrsim1$ for the phase plate to be effective.

In the case of radiation-sensitive samples, a finite beam can offer the advantage of avoiding sample exposure, and thus damage, outside the field of view. It can also be argued that, in reality, beams do have a finite width. Notwithstanding these remarks, if the incident beam has finite width, then a phase plate in Fourier space constitutes \emph{only an approximation} to the Zernike phase condition. As we have shown, a good choice of $R$ and $K$ leads to a highly effective approximation, and a poor choice renders the approximation less effective (as in Ref.~\cite{VegaIbanezVerbeeck2025}).

We showed that, for an arbitrary incident state, a Zernike operator can be defined such that the Zernike phase condition is achieved exactly. This includes the finite beam just discussed, and any other case. The Zernike operator has a simple mathematical form, and the formalism makes clear the necessity of detecting the electrons in real space where the sample's potential has a representation which is diagonal and real. Finding an implementation of the Zernike operator in terms of optical elements is challenging in the general case. 

Inspired by the existence of the Zernike operator, we have considered an approximate optical implementation of Zernike speckle imaging, where the incident wave is a speckle wave. In previous work \cite{DwyerPaganin2024}, we found that such an incident wave optimizes 4D STEM, in the sense that it can then achieve about half of the quantum limit across a large range of spatial frequencies. Here, we found that, for correctly chosen parameters $K$ and $R$, our approximate implementation of Zernike speckle imaging can achieve $>95$\% of the quantum limit (within our stated idealizations). An incident speckle wave constitutes a significant departure from the plane wave, or plane wave-like, illumination typically associated with Zernike phase-contrast imaging. Hence the existence of such an implementation underscores the potential generality of the Zernike method.

As a final remark, we wish to remind the reader that the Fisher information, and the corresponding Cramer-Rao lower bound on the parameter variance (which so far we have not mentioned in this work), apply in the \emph{asymptotic} regime where the statistics become Gaussian. Thus, it may be that the results of this work may not apply at extremely low doses. If so, it will be necessary to employ a different lower bound, such as the quantum Ziv-Zakai bound \cite{Ziv-Zakai1969, Tsang2012}. We refer to our previous work \cite{Dwyer2023} for further details.

\bibliography{generalizedZernike.bib}

\newpage

\begin{table}[h!]
\caption{Parameters for the Gaussian atomic potential.}
\begin{center}
\begin{tabular}{|c||c|}\hline
$N_x$, $N_y$ & 2048, 2048 \\
$A_j$        & 0.1 \\
FWHM         & 8.0 \\
$(x_j, y_j)$ & $(R/2, 0)$ \\\hline
\end{tabular}
\end{center}
\label{tab:atomic pos parameters}
\end{table}%

\begin{table}[h!]
\caption{Fisher information on atomic position $x_j$ (or $y_j$) from finite beams, as a function of beam radius $R$ and phase plate cutoff $K$, quoted as a fraction of the quantum limit.}
\begin{center}
\begin{tabular}{|c||cccccccc|}\hline
\backslashbox{$R$}{$K$} & $\dfrac{2}{2048}$ & $\dfrac{4}{2048}$ & $\dfrac{8}{2048}$ & $\dfrac{16}{2048}$ & $\dfrac{32}{2048}$ & $\dfrac{64}{2048}$ & $\dfrac{128}{2048}$ & $\dfrac{256}{2048}$ \\ \hline\hline
128 & 0.01      & 0.40      & {\bf0.99} & {\bf0.99} & {\bf0.99} & 0.92 & 0.46 & 0.01 \\
256 & 0.27      & {\bf1.00} & {\bf0.99} & {\bf1.00} & {\bf0.99} & 0.92 & 0.46 & 0.01 \\
512 & {\bf1.00} & {\bf0.99} & {\bf1.00} & {\bf1.00} & {\bf0.99} & 0.92 & 0.47 & 0.01 \\\hline
\end{tabular}
\end{center}
\label{tab:atomic pos}
\end{table}%

\begin{figure}[h!] 
\includegraphics[width=0.405\columnwidth]{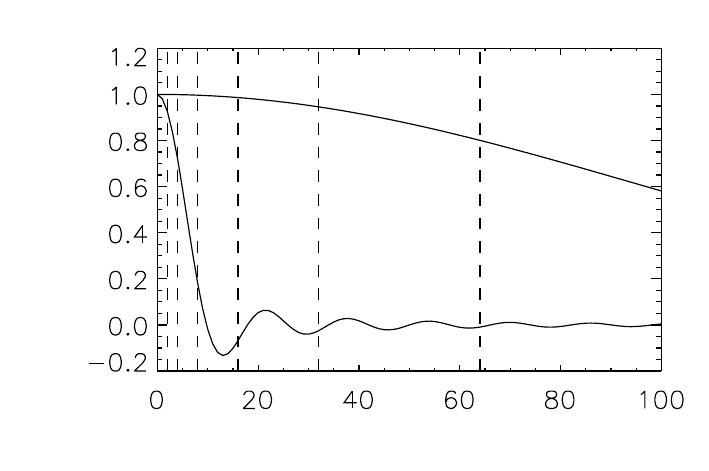}\\\vspace{-5mm}
\includegraphics[width=0.405\columnwidth]{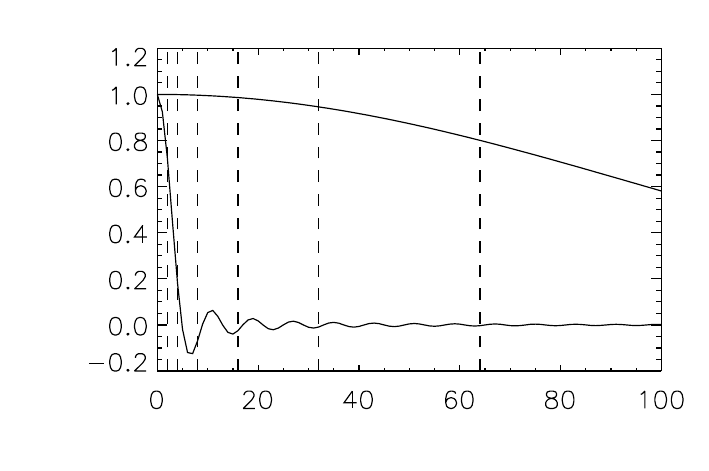}\\\vspace{-5mm}
\includegraphics[width=0.405\columnwidth]{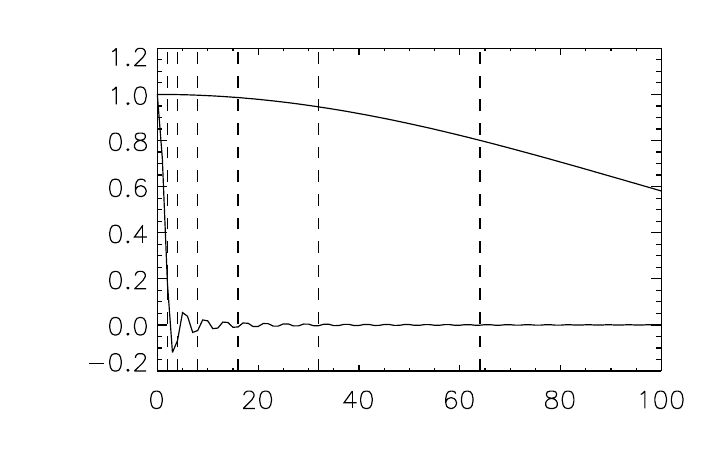}
\caption{\label{fig:atomic pos ZPP} Plots of $\tilde\psi_0(\k)$ vs $|\k|$ for beam radii $R=128$ (top), $R=256$ (middle) and $R=512$ (bottom). Vertical dashed lines indicate some values of the phase plate cutoff $K$ (abscissa values are multiplied by $2048$). Each plot also shows the potential $\tilde V(\k)$ relevant to the atomic position measurement. The potential $\tilde V(\k)$ relevant to the Fourier coefficient measurement is a pair of delta functions at $\pm|\g| = \pm 128$ (not shown).}
\end{figure}

\newpage

\begin{table}[h!]
\caption{Parameters for the Fourier coefficient $V_\g$.}
\begin{center}
\begin{tabular}{|c||c|}\hline
$N_x$, $N_y$ & 2048, 2048 \\
$d$          & 16 \\
$|\g|$       & $128/2048$\\
$|V_\g|$     & 0.05 \\
$\phi_\g$    & 0 \\\hline
\end{tabular}
\end{center}
\label{tab:Fourier coeff parameters}
\end{table}%

\begin{table}[h!]
\caption{Fisher information on Fourier modulus from finite beams, as a function of beam radius $R$ and phase plate cutoff $K$, quoted as a fraction of the quantum limit.}
\begin{center}
\begin{tabular}{|c||cccccccc|}\hline
\backslashbox{$R$}{$K$} & $\dfrac{2}{2048}$ & $\dfrac{4}{2048}$ & $\dfrac{8}{2048}$ & $\dfrac{16}{2048}$ & $\dfrac{32}{2048}$ & $\dfrac{64}{2048}$ & $\dfrac{128}{2048}$ & $\dfrac{256}{2048}$ \\ \hline\hline
128 & 0.01 & 0.35 & 0.84      & 0.92      & {\bf0.96} & {\bf0.98} & 0.41 & 0.00 \\
256 & 0.24 & 0.85 & 0.92      & {\bf0.96} & {\bf0.98} & {\bf0.99} & 0.46 & 0.00 \\
512 & 0.86 & 0.92 & {\bf0.96} & {\bf0.98} & {\bf0.99} & {\bf0.99} & 0.58 & 0.00 \\\hline
\end{tabular}
\end{center}
\label{tab:Fourier modulus}
\end{table}%

\begin{table}[h!]
\caption{Fisher information on Fourier phase from a finite beams, as a function of beam radius $R$ and phase plate cutoff $K$, quoted as a fraction of the quantum limit.}
\begin{center}
\begin{tabular}{|c||cccccccc|}\hline
\backslashbox{$R$}{$K$} & $\dfrac{2}{2048}$ & $\dfrac{4}{2048}$ & $\dfrac{8}{2048}$ & $\dfrac{16}{2048}$ & $\dfrac{32}{2048}$ & $\dfrac{64}{2048}$ & $\dfrac{128}{2048}$ & $\dfrac{256}{2048}$ \\ \hline\hline
128 & 0.01 & 0.35 & 0.84      & 0.92      & {\bf0.96} & {\bf0.97} & 0.40 & 0.00 \\
256 & 0.24 & 0.85 & 0.92      & {\bf0.96} & {\bf0.98} & {\bf0.99} & 0.46 & 0.00 \\
512 & 0.86 & 0.92 & {\bf0.96} & {\bf0.98} & {\bf0.99} & {\bf1.00} & 0.58 & 0.00 \\\hline
\end{tabular}
\end{center}
\label{tab:Fourier phase}
\end{table}%

\newpage

\newpage

\newpage

\newpage

\newpage

\newpage 

\begin{table}[h!]
\caption{Fisher information on Fourier modulus and phase from a speckle beam with $K=128/2048$, as a function of phase plate cutoff $R$, quoted as a fraction of the quantum limit.}
\begin{center}
\begin{tabular}{|c||cccccccc|}\hline
$R$ & 2 & 4 & 8 & 16 & 32 & 64 & 128 & 256 \\
$F_{\mu\mu}$ modulus & 0.02 & 0.35 & 0.88     & 0.94      & {\bf0.97} & {\bf0.98} & {\bf0.98} & 0.94 \\
$F_{\mu\mu}$ phase   & 0.02 & 0.35 & 0.89      & 0.94      & {\bf0.97} & {\bf0.98} & {\bf0.98} & 0.94 \\\hline
\end{tabular}
\end{center}
\label{tab:speckle}
\end{table}%

\end{document}